\documentclass{vldb}

\usepackage[caption=false]{subfig}  
\usepackage{graphicx}
\usepackage[usenames]{color}

\usepackage{multicol}
\usepackage{syntax}
\usepackage{booktabs}
\usepackage[normalem]{ulem}

\usepackage{latexsym}
\usepackage{stmaryrd}

\usepackage{multirow}

\usepackage{pstricks}
\usepackage{pst-node}
\usepackage{pst-tree}

\usepackage{paralist}
\usepackage{url}

\usepackage{algorithmic}

\newtheorem{proposition}{Proposition}

\newdef{definition}{Definition} 
\newdef{example}{Example} 
\newdef{algorithm}{Algorithm}

\newcommand{\T}{\mathcal{T}}
\newcommand{\U}{\mathcal{U}}
\newcommand{\A}{\mathcal{A}}
\newcommand{\B}{\mathcal{B}}
\newcommand{\C}{\mathcal{C}}

\newcommand{\D}{\mathbf{D}}

\renewcommand{\phi}{\varphi}
\newcommand{\tuple}[1]{{\langle#1\rangle}}

\newcommand{\btuple}[1]{{\langle\textcolor{blue}{#1}\rangle}}
\newcommand{\bttuple}[1]{{\langle\textcolor{blue}{\mathrm{#1}}\rangle}}
\newcommand{\data}[2]{\langle\textcolor{blue}{{#1}\!:\!{#2}}\rangle}

\newcommand{\anc}{\mathrm{anc}}
\newcommand{\rel}{\mathrm{rel}}

\def\punto{$\hspace*{\fill}\Box$}
\newcommand{\cpar}[1]{ {\noindent\bf #1} }

\newcommand{\jakub}[1]{\textcolor{PineGreen}{{\bf jakub:} #1}}
\newcommand{\new}[1]{\textcolor{magenta}{#1}}

\newcommand{\nop}[1]{}

\begin{document}

\title{FDB: A Query Engine for Factorised Relational Databases}

\numberofauthors{1}

\author{
\alignauthor
Nurzhan Bakibayev, Dan Olteanu, and Jakub Z\'{a}vodn\'{y}\\
       \affaddr{Department of Computer Science, University of Oxford, OX1 3QD, UK}
       \email{\{nurzhan.bakibayev, dan.olteanu, jakub.zavodny\}@cs.ox.ac.uk}
}

\maketitle
\begin{abstract}
  Factorised databases are relational databases that use compact
  factorised representations at the physical layer to reduce data
  redundancy and boost query performance. 

  This paper introduces FDB, an in-memory query engine for
  select-project-join queries on factorised databases. Key components
  of FDB are novel algorithms for query optimisation and evaluation
  that exploit the succinctness brought by data factorisation.
  Experiments show that for data sets with many-to-many relationships
  FDB can outperform relational engines by orders of magnitude.
\end{abstract}

\nop{


}


\section{Introduction}

This paper introduces FDB, an in-memory query engine for
select-project-join queries on factorised relational data.

At the outset of this work lies the observation that relations can
admit compact, factorised representations that can effectively boost
the performance of relational processing. The relationship between
relations and their factorised representations is on a par with the
relationship between logic functions in disjunctive normal form and
their equivalent nested forms obtained by algebraic factorisation.

\begin{figure*}[t]
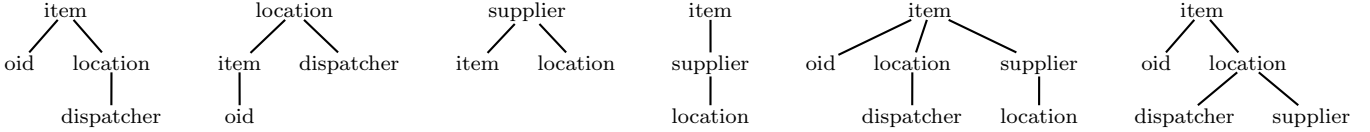

\begin{small}
\[
\psset{levelsep=7mm, nodesep=1pt, treesep=5mm}
\pstree{\TR{\textrm{item}}}
{
  \TR{\textrm{oid}}
  \pstree{\TR{\textrm{location}}}
  {
    \TR{\textrm{dispatcher}}
  }
}%
\hspace*{2.5em}
\psset{levelsep=7mm, nodesep=1pt, treesep=5mm}
\pstree{\TR{\textrm{location}}}
{
  \pstree{\TR{\textrm{item}}}
  {
    \TR{\textrm{oid}}
  }
  \TR{\textrm{dispatcher}}
}%
\hspace*{2.5em}
\psset{levelsep=7mm, nodesep=1pt, treesep=5mm}
\pstree{\TR{\textrm{supplier}}}
{
  \TR{\textrm{item}}
  \TR{\textrm{location}}
}%
\hspace*{2.5em}
\psset{levelsep=7mm, nodesep=1pt, treesep=5mm}
\pstree{\TR{\textrm{item}}}
{
  \pstree{\TR{\textrm{supplier}}}
  {
    \TR{\textrm{location}}
  }
}%
\hspace*{2.5em}
\psset{levelsep=7mm, nodesep=1pt, treesep=5mm}
\pstree{\TR{\textrm{item}}}
{
  \TR{\textrm{oid}}
  \pstree{\TR{\textrm{location}}}
  {
    \TR{\textrm{dispatcher}}
  }
  \pstree{\TR{\textrm{supplier}}}
  {
    \TR{\textrm{location}}
  }
}%
\hspace*{2.5em}
\psset{levelsep=7mm, nodesep=1pt, treesep=5mm}
\pstree{\TR{\textrm{item}}}
{
  \TR{\textrm{oid}}
  \pstree{\TR{\textrm{location}}}
  {
    \TR{\textrm{dispatcher}}
    \TR{\textrm{supplier}}
  }
}%
\]
\end{small}

\vspace*{-1em}
\caption{Factorisation trees used in Example~\ref{ex:running}. From
  left to right: $\T_1$ and $\T_2$ for the result of query $Q_1$;
  $\T_3$ and $\T_4$ for the result of $Q_2$; $\T_5$ is obtained after
  joining $\T_1$ and $\T_4$ on item, and $\T_6$ is $\T_5$ after
  joining on location.}
\label{fig:ex-ftrees}
\vspace*{-1em}
\end{figure*}

\begin{example}\label{ex:running}
  Consider a database of a grocery retailer containing delivery
  orders, stock availability at different locations, availability of
  dispatcher units for the individual locations, and grocery producers
  with items they produce and locations they supply to
  (Figure~\ref{fig:ex-db}).  A query $Q_1$ that finds all orders with
  their respective items, possible locations to retrieve them from,
  and dispatchers available to deliver them, returns the following
  result (shown only partially):

\begin{center}
\vspace*{-.5em}
\begin{scriptsize}
\begin{tabular}{llll}
   \multicolumn{4}{c}{$Q_1 = \textrm{Order}\Join_{\textrm{item}}\textrm{Store}\Join_{\textrm{location}}\textrm{Disp}$}\\\toprule
oid & item & location & dispatcher \\\midrule
 01 & Milk & Istanbul & Adnan \\
 01 & Milk & Istanbul & Yasemin \\
 01 & Milk & Izmir & Adnan \\
 01 & Milk & Antalya & Volkan \\
\nop{ 01 & Cheese & Istanbul & Adnan \\
 01 & Cheese & Istanbul & Yasemin\\}
 \multicolumn{4}{c}{$\dots$}\\\bottomrule
\end{tabular}
\end{scriptsize}
\vspace*{-.5em}
\end{center}

This query result can be expressed as a relational expression built
using singleton relations, union, and product, whereby each singleton
relation $\bttuple{v}$ holds one value \texttt{v}, each tuple is a
product of singleton relations, and an arbitrary relation is a union of
products of singleton relations:
\begin{scriptsize}
\begin{align*}
&\bttuple{01}\times\bttuple{Milk}\times\bttuple{Istanbul}\times\bttuple{Adnan}\cup\\
&\bttuple{01}\times\bttuple{Milk}\times\bttuple{Istanbul}\times\bttuple{Yasemin}\cup\\
&\bttuple{01}\times\bttuple{Milk}\times\bttuple{Izmir}\times\bttuple{Adnan}\cup\\
&\bttuple{01}\times\bttuple{Milk}\times\bttuple{Antalya}\times\bttuple{Volkan}\cup \nop{\\
&\bttuple{01}\times\bttuple{Cheese}\times\bttuple{Istanbul}\times\bttuple{Adnan}\cup\\
&\bttuple{01}\times\bttuple{Cheese}\times\bttuple{Istanbul}\times\bttuple{Yasemin}\cup} \ldots
\end{align*}
\end{scriptsize}
A more compact equivalent representation can be obtained by algebraic
factorisation using distributivity of product over union and
commutativity of product and union:
\begin{scriptsize}
\begin{align*}
 &\bttuple{Milk}\times\bttuple{01}\times(\bttuple{Istanbul}\times(\bttuple{Adnan} \cup \bttuple{Yasemin}) \cup \\
 &\phantom{\bttuple{Milk}\times\bttuple{01}\times(}\bttuple{Izmir}\times\bttuple{Adnan} \cup \bttuple{Antalya}\times\bttuple{Volkan}) \cup \\
 &\bttuple{Cheese}\times(\bttuple{01} \cup \bttuple{03})\times(\bttuple{Istanbul}\times(\bttuple{Adnan} \cup \bttuple{Yasemin}) \cup \\
 &\phantom{\bttuple{Cheese}\times(\bttuple{01} \cup \bttuple{03})\times(}\bttuple{Antalya}\times\bttuple{Volkan}) \cup \\
 &\bttuple{Melon}\times(\bttuple{02} \cup \bttuple{03})\times\bttuple{Istanbul}\times(\bttuple{Adnan} \cup \bttuple{Yasemin})
\end{align*}
\end{scriptsize}
\nop{We can read it as ``Milk for order \#01 can be delivered from
  Istanbul by Adnan or Yasemin, or from Izmir by Adnan, or from
  Antalya by Volkan,'' and so on.} This {\em factorised
  representation} has the following structure: for each item, we
construct a union of its possible orders and a union of its possible
locations with dispatchers. This nesting structure together with the
attribute names form the schema of the factorised representation,
which we call a {\em factorisation tree}, or f-tree for short.

Figure~\ref{fig:ex-ftrees} depicts several f-trees; the leftmost one
($\T_1$) captures the nesting structure of the above factorisation.
The second f-tree ($\T_2$) is an alternative nesting structure for the
same query result, where for each location, we construct a union of
its items and orders and a union of dispatchers:
\begin{scriptsize}
\begin{align*}
 &\bttuple{Istanbul}\times(\bttuple{Milk}\times\bttuple{01} \cup \bttuple{Cheese}\times(\bttuple{01} \cup \bttuple{03}) \cup \\
 &\phantom{\bttuple{Istanbul}\times(}\bttuple{Melon}\times(\bttuple{02} \cup \bttuple{03}))\times(\bttuple{Adnan} \cup \bttuple{Yasemin}) \cup \\
 &\bttuple{Izmir}\times\bttuple{Milk}\times\bttuple{01}\times\bttuple{Adnan} \cup \\
 &\bttuple{Antalya}\times(\bttuple{Milk}\times\bttuple{01} \cup \bttuple{Cheese}\times(\bttuple{01} \cup \bttuple{03}))\times\bttuple{Volkan}
\end{align*}
\end{scriptsize}
The factorised result of the query
$Q_2=\textrm{Produce}\Join_{\textrm{supplier}}\textrm{Serve}$ over the
f-tree $\T_3$ given in Figure~\ref{fig:ex-ftrees} is:
\begin{scriptsize}
\begin{align*}
 &\bttuple{Guney}\times(\bttuple{Milk} \cup \bttuple{Cheese})\times\bttuple{Antalya} \cup \\
 &\bttuple{Dikici}\times\bttuple{Milk}\times(\bttuple{Istanbul} \cup \bttuple{Izmir} \cup \bttuple{Antalya}) \cup \\
 &\bttuple{Byzantium}\times\bttuple{Melon}\times\bttuple{Istanbul}\hspace*{8em}\Box
\end{align*}
\end{scriptsize}
\end{example}


\begin{figure*}[t]
\begin{center}
\begin{scriptsize}
\subfloat{
\begin{tabular}{@{~}l@{~}l@{~}}
\multicolumn{2}{c}{\textrm{Orders}}\\\toprule
oid & item \\\midrule
 01 & Milk \\
 01 & Cheese \\
 02 & Melon \\
 03 & Cheese \\
 03 & Melon\\\bottomrule
\ \\
\end{tabular}
}\quad
\subfloat{
\begin{tabular}{@{~}l@{~}l@{~}l@{~}}
\multicolumn{2}{c}{\textrm{Store}}\\\toprule
 location & item\\\midrule
 Istanbul & Milk \\
 Istanbul & Cheese \\
 Istanbul & Melon \\
 Izmir & Milk \\
 Antalya & Milk \\
 Antalya & Cheese\\\bottomrule
\end{tabular}
}\quad
\subfloat{
\begin{tabular}{@{}l@{~}@{~}l@{~}@{~}}
\multicolumn{2}{c}{\textrm{Disp}}\\\toprule
 dispatcher & location \\\midrule
 Adnan & Istanbul \\
 Adnan & Izmir \\
 Yasemin & Istanbul \\
 Volkan & Antalya\\\bottomrule
\ \\
\ \\
\end{tabular}
}\quad
\subfloat{
\begin{tabular}{@{~}l@{~}l@{~}}
\multicolumn{2}{c}{\textrm{Produce}}\\\toprule
 supplier & item \\\midrule
 Guney & Milk \\
 Guney & Cheese \\
 Dikici & Milk \\
 Byzantium & Melon\\\bottomrule
\ \\
\ \\
\end{tabular}
}\quad
\subfloat{
\begin{tabular}{@{~}l@{~}l@{~}l@{~}}
\multicolumn{2}{c}{\textrm{Serve}}\\\toprule
 supplier & location \\\midrule
 Guney & Antalya \\
 Dikici & Istanbul \\
 Dikici & Izmir \\
 Dikici & Antalya \\
 Byzantium & Istanbul\\\bottomrule
\ \\
\end{tabular}
}
\end{scriptsize}
\end{center}
\vspace*{-1em}
\caption{An example database for a grocery retailer.}
\label{fig:ex-db}
\vspace*{-1em}
\end{figure*}



Factorisations are ubiquitous. They are arguably most known for
minimisation of Boolean functions~\cite{brayton87} but can be useful
in a number of read-optimised database scenarios. The scenario we
consider in this paper is that of factorising large intermediate and
final results to speed-up query evaluation on data sets with
many-to-many relationships. A further scenario we envisage is that of
compiled databases: these are static databases, such as databases
encoding the human genome~\cite{GDB1998}, that can be aggressively
factorised to efficiently support a particular scientific workload. In
provenance and probabilistic databases, factorisations of provenance
polynomials~\cite{Green:PODS:2007} are used for compact encoding of
large provenance (the GeneOntology database has records with
10MB provenance)~\cite{OZ11b} and for efficient query
evaluation~\cite{OH2008,PDB-BOOK11}.  Factorisations are a natural fit
whenever we deal with a large space of possibilities or choices. For
instance, data models for design specifications, such as the AND/OR
trees~\cite{INV1991}, are based on incompleteness and non-determinism
and are captured by factorised representations.  Formalisms for
incomplete information, such as world-set
decompositions~\cite{AKO07WSD,OKA08gWSD}, rely on factorisations of
universal relations encoding very large sets of possible worlds; they
are products of unions of products of tuples. Outside data management
scenarios, factorised relations can be used to compactly represent the
space of feasible solutions to configuration problems in constraint
satisfaction, where we need to connect a fixed finite set of given
components so as to meet a given objective while respecting given
constraints~\cite{Jeavons11}.


Factorised representations have several key properties that make them
appealing in the above mentioned scenarios.

They {\em can be exponentially more succinct} than the relations they
encode. For instance, a product of $n$ relations needs size
exponential in $n$ for a relational result, but only linear in the
size of the input relations for a factorised result. Recent work has
established tight bounds on the size of factorised query
results~\cite{OZ12}: For any select-project-join query $Q$, there is a
rational number $s(Q)$ such that for any database $\D$, there exists a
factorised representation $E$ of $Q(\D)$ with size $O(|\D|^{s(Q)})$,
and within the class of representations whose structures are given by
factorisation trees, there is no factorisation of smaller size. The
parameter $s(Q)$ is the fractional edge cover number of a particular
subquery of $Q$, and there are arbitrarily large queries $Q$ for which
$s(Q)=1$. Moreover, the exponential gap between the sizes of $E$ and
of $Q(\D)$ also holds between the times needed to compute $E$ and
$Q(\D)$ directly from the input database $\D$.

Further succinctness can be achieved using dictiona\-ry-based
compression and null suppression of data
values~\cite{CompressionOracle03}. Compressing entire vertical
partitions of relations as done in c-store~\cite{Abadi:Compression06}
is not compatible with our factorisation approach since it breaks the
relational structure.

Notwithstanding succinctness, factorised representations of query
results {\em allow for fast (constant-delay) enumeration of tuples}.
More succinct representations are definitely possible, e.g., binary
join decompositions~\cite{Gottlob11} or just the pair of the query and
the database~\cite{Durand07}, but then retrieving any tuple in the
query result is already NP-hard. Factorised representations can thus
be seen as compilations of query results that allow for efficient
subsequent processing.

By construction, factorised representations {\em reduce redundancy in
  the data and boost query performance} using a mixture of vertical
(product) and horizontal (union) data partitioning. This goal is
shared with a large body of work on normal forms~\cite{AHV95} and
columnar stores~\cite{MonetDB99} that considers join (or general
vertical) decompositions, and with partitioning-based automated
physical database design~\cite{Agrawal:Partitioning04,Hyrise2010}. In
the latter case, the focus is on partitioning input data such that the
performance of a particular workload is maximised.

Finally, factorised representations {\em are relational algebra
expressions} with well-understood semantics. Their relational nature
sets them apart from XML documents, object-oriented data\-ba\-ses, and
nested objects~\cite{AHV95}, where the goal is to avoid the rigidity
of the relational model. Moreover, in our setting, a query result can
admit several equivalent factorised representations and the goal is to
find one of small size. The Verso project~\cite{Verso1986} points out
compactness and modelling benefits of non-first-normal-form relations
and considers hierarchical data representations that are special cases
of factorised representations. It does not focus on factorisations and
thus neither on the search for ones of small sizes.


A factorised database presents relations at the logical layer but uses
succinct factorised representations at the physical layer. The FDB
query engine can thus not only compute factorised query results for
input relational databases, but can evaluate queries directly on input
factorised databases.

\begin{example}\label{ex:running-continued}
  Consider now the query $Q_1\Join_{\textrm{location,item}}Q_2$ on
  factorised representations: Find possible suppliers of ordered
  items. Joining the above factorisations over the f-trees $\T_1$ and
  $\T_3$ on the attributes $\textrm{location}$ and $\textrm{item}$ is
  not immediate, since tuples with equal values for
  $\textrm{location}$ and $\textrm{item}$ appear scattered in the
  factorisation over $\T_3$.  If we restructure the factorisation of
  $Q_2$'s result to follow the f-tree $\T_4$ so that tuples are
  grouped by $\textrm{item}$ first, we obtain\vspace*{-.5em}

\begin{scriptsize}
\begin{align*}
 &\bttuple{Milk}\times(\bttuple{Guney}\times\bttuple{Antalya} \cup \\
 &\phantom{\bttuple{Milk}\times(}\bttuple{Dikici}\times(\bttuple{Istanbul} \cup \bttuple{Izmir} \cup \bttuple{Antalya})) \\
 &\bttuple{Cheese}\times\bttuple{Guney}\times\bttuple{Antalya} \cup \\
 &\bttuple{Melon}\times\bttuple{Byzantium}\times\bttuple{Istanbul},
\end{align*}
\end{scriptsize} 
which can be readily joined with the factorisation over $\T_1$ on the
attribute item, since both factorisations have items as topmost
values. The factorisation of the join on item follows the f-tree
$\T_5$, where we simply merged the roots of the two f-trees. An
excerpt of this factorisation is
\begin{scriptsize}
\begin{align*}
 &\bttuple{Milk}\times\bttuple{01}\times(\bttuple{Istanbul}\times(\bttuple{Adnan} \cup \bttuple{Yasemin}) \cup \\
 &\phantom{\bttuple{Milk}\times\bttuple{01}\times(}\bttuple{Izmir}\times\bttuple{Adnan} \cup \bttuple{Antalya}\times\bttuple{Volkan}) \\
 &\phantom{\bttuple{Milk}}\times(\bttuple{Guney}\times\bttuple{Antalya} \cup \\
 &\phantom{\bttuple{Milk}\times(}\bttuple{Dikici}\times(\bttuple{Istanbul} \cup \bttuple{Izmir} \cup \bttuple{Antalya})) \cup\dots,
\end{align*}
\end{scriptsize}
To perform the second join condition on location, we first need for
each item to rearrange the subexpression for suppliers and locations,
so that it is grouped by locations as opposed to suppliers. This
amounts to swapping supplier and location in $\T_5$. The join on
location can now be performed between the possible locations of each
item. The obtained factorisation follows the schema $\T_6$ in
Figure~\ref{fig:ex-ftrees}.
\nop{
\begin{small}
\begin{align*}
 &\bttuple{Milk}\bttuple{01}(\bttuple{Istanbul}(\bttuple{Adnan} \cup \bttuple{Yasemin})\bttuple{Dikici} \cup \\
 &\phantom{\bttuple{Milk}\bttuple{01}(}\bttuple{Izmir}\bttuple{Adnan}\bttuple{Dikici} \cup \\
 &\phantom{\bttuple{Milk}\bttuple{01}(} \bttuple{Antalya}\bttuple{Volkan}(\bttuple{Dikici} \cup \bttuple{Guney})) \cup \\
 &\dots,
\end{align*}
\end{small}
}
\punto
\end{example}

Examples~\ref{ex:running} and~\ref{ex:running-continued} highlight
challenges involved in computing factorised representations of query
results.

Firstly, a query result may have different (albeit equivalent)
factorised representations whose sizes can differ by an exponential
factor. We seek f-trees that define succinct representations of query
results for all input (relational or factorised) databases. Such
f-trees can be statically derived from the query and the input schema,
but are independent of the database content. Query optimisation thus
has to consider two objectives: minimising the cost of computing a
factorised query result from the (possibly factorised) input database,
and minimising the size of this output representation.  In addition to
the standard query operators selection, projection, and product, the
search space for a good query and factorisation plan, or f-plan for
short, needs to consider specific operators for restructuring schemas
and factorisations. We propose two such operators: a swap operator,
which exchanges a given child with its parent in an f-tree, and a
push-up operator, which moves an entire sub-tree up in the f-tree.
For instance, the swap operator is used to transform the f-tree $\T_3$
into $\T_4$ in Figure~\ref{fig:ex-ftrees}.  The selection operator is
used to merge the item nodes in the f-trees $\T_1$ and $\T_4$ and
create the f-tree $\T_5$. The transformation of $\T_5$ into $\T_6$,
which corresponds to a join on location, needs a swap of supplier and
location and a merge of the two location nodes.

Secondly, we would like to compute the factorised query result as
efficiently as possible. This means in particular that we must avoid
the computation of intermediate results in relational, un-factorised
form.  Our query engine has algorithms for each operator selection,
projection, product, swap, and push-up. These algorithms use time
(quasi)linear in the sizes of input and output representations and
ensure that the f-tree of the resulting factorisation is optimal with
respect to tight size bounds that can be derived from the input
f-tree and the operator.

The main contributions of this paper are as follows:
\begin{itemize}
\item We address new challenges to query optimisation in the presence
  of factorised data and restructuring operators. In addition to the
  cost of computing the factorised query result, we also need to
  consider the size of the resulting factorisation.

  We give exhaustive and heuristic optimisation algorithms for
  computing f-plans whose outcomes are factorised query results. As
  cost metric, we use selectivity and cardinality estimates and a
  parameter that defines tight bounds on the sizes of the factorised
  result and of the temporary results.

\item We give algorithms for the evaluation of each f-plan operator on
  factorised data. They are optimal with respect to time complexity
  and to tight size bounds inferred from the input f-tree and the
  operator.

\item The optimisation and evaluation algorithms have been
  implemented in the FDB in-memory query engine.

\item We report on an extensive experimental evaluation showing that 
FDB can outperform a homebred in-memory and two open-source (SQLite
and PostgreSQL) relational query engines by orders of magnitude.
\end{itemize}

\nop{
\cpar{Related work.} 
Data compression shares with factorisation the goal of compact data
representation, as used e.g., for column
compression~\cite{SAB+2005,Hyrise2010} and value compression in
Oracle~\cite{CompressionOracle03}, and the XMill compressor of XML
data~\cite{Liefke:CompressedXML:2000}. Structure-preserving
compression of trees into directed acyclic graphs has been used to
boost the performance of XML query
evaluation~\cite{CK:CompressedXML:2003}.

Finally, we note the connection to query tractability in relational
databases, e.g.,~\cite{AHV95,Gottlob09a}. The bulk of tractability
characterisations is for Boolean queries. Non-Boolean queries such as
the product of $n$ relations require an evaluation time exponential in
$n$, yet have factorised representations of size and computable in
time linear in the input.
}


\nop{
\jakub{Alternative depiction of the first factorisation is given below, but similar problems arise as in Maria's project: we need to draw the delimiting lines carefully, to show associativity of combinations.
\begin{center}
\vspace{-2pt}
\begin{small}
\begin{tabular}{@{~}c@{~}c@{~}c@{~}c@{~}}
   \multicolumn{4}{c}{$\textrm{Order}\Join_{\textrm{item}}\textrm{Store}\Join_{\textrm{location}}\textrm{Emp}$}\\\hline
 item & id & location & dispatcher \\   
\hline
 \multirow{4}{*}{Milk} & \multirow{4}{*}{01} & \multirow{2}{*}{Istanbul} & Adnan \\ \cline{4-4}
  & & & Yasemin \\
\cline{3-4}
  & & Izmir & Adnan \\
\cline{3-4}
  & & Antalya & Volkan \\
\cline{1-4}
 \multirow{3}{*}{Cheese} & \multirow{3}{*}{\begin{tabular}{c}01 \\ \hline 03\end{tabular}} & \multirow{2}{*}{Istanbul} & Adnan \\
\cline{4-4}
  & & & Yasemin \\
\cline{3-4}
  & & Antalya & Volkan
\end{tabular}
\\[3pt]
\dots
\end{small}
\vspace{-2pt}
\end{center}
}
}


\section{F-Representations and F-Trees}
\label{sec:frep}

We next recall the notions of factorised representations and
factorisation trees, as well as results on tight size bounds for such
factorised representations over factorisation trees~\cite{OZ12}.
\medskip

{\noindent\bf Factorised representations} of relations are algebraic
expressions constructed using singleton relations and the relational
operators union and product.

\begin{definition}
  A {\em factorised representation} $E$, or f-represen\-tation for
  short, over a set $\mathcal{S}$ of attributes and domain
  $\mathcal{D}$ is a relational algebra expression of the form
\begin{compactitem}
\item $\emptyset$, the empty relation over schema $\mathcal{S}$;
\item $\tuple{}$, the relation consisting of the nullary tuple, if
  $\mathcal{S}=\emptyset$;
\item $\data{A}{a}$, the unary relation with a single tuple with value
  $a$, if $\mathcal{S}=\{A\}$ and $a$ is a value in the domain
  $\mathcal{D}$;
\item $(E)$, where $E$ is an f-representation over $\mathcal{S}$;
\item $E_1 \cup \dots \cup E_n$, where each $E_i$ is an
  f-representation over~$\mathcal{S}$;
\item $E_1 \times \dots \times E_n$, where each $E_i$ is an
  f-representation over~$\mathcal{S}_i$ and $\mathcal{S}$ is the
  disjoint union of all $\mathcal{S}_i$.
\end{compactitem}
\end{definition}

An expression $\data{A}{a}$ is called an \emph{A-singleton} and the
expression $\tuple{}$ is called the nullary singleton. The
\emph{size} $|E|$ of an f-representation $E$ is the number of
singletons in $E$.

Any f-representation over a set $\mathcal{S}$ of attributes can be
interpreted as a database over schema $\mathcal{S}$.
Example~\ref{ex:running} gives several f-representations, w\-he\-re
singleton types are dropped for compactness reasons. For instance,
$\bttuple{Istanbul}$ $\times(\bttuple{Adnan} \cup \bttuple{Yasemin})$
represents a relation with schema $\{$location, dis\-patcher$\}$ and
tuples
$(\textcolor{blue}{\mathrm{Istanbul}},\textcolor{blue}{\mathrm{Adnan}}),
(\textcolor{blue}{\mathrm{Istanbul}},\textcolor{blue}{\mathrm{Yasemin}})$.

F-representations form a representation system for relational
databases. It is {\em complete} in the sense that any databa\-se can
be represented in this system, but not injective since there exist
different f-representations for the same database. The space of
f-representations of a database is defined by the distributivity of
product ($\times$) over union ($\cup$). Under the RAM model with
uniform cost measure, the tuples of a given f-representation $E$ over
a set $\mathcal{S}$ of attributes can be enumerated with $O(|E|)$
space and precomputation time, and $O(|\mathcal{S}|)$ delay between
successive tuples.
\medskip

{\noindent\bf Factorisation trees} define classes of
f-repre\-sentations over a set of attributes and with the same nesting
structure.

\begin{definition}
  A \emph{factorisation tree}, or f-tree for short, over a schema
  ${\cal S}$ of attributes is an unordered rooted forest with each
  node labelled by a non-empty subset of ${\cal S}$ such that each
  attribute of ${\cal S}$ labels exactly one node.
\end{definition}

Given an f-tree $\T$, an f-representation over $\T$ is recursively
defined as follows:

\begin{compactitem}
\item If $\T$ is a forest of trees $\T_1,\dots,\T_k$, then
  \[E = E_1 \times \dots \times E_k\] where each $E_i$ is an
  f-representation over $\T_i$.

\item If $\T$ is a single tree with a root labelled by
  $\{A_1,\dots,A_k\}$ and a non-empty forest $\U$ of children, then
  \[E = \textstyle\bigcup_a
  \data{A_1}{a}\times\dots\times\data{A_k}{a}\times E_a\] where each
  $E_a$ is an f-representation over $\U$ and the union $\bigcup_a$ is
  over a collection of distinct values $a$.

\item If $\T$ is a single node labelled by $\{A_1,\dots,A_k\}$, then
\[E = \textstyle\bigcup_a \data{A_1}{a}\times\dots\times\data{A_k}{a}.\]

\item If $\T$ is empty, then $E = \emptyset$ or $E = \tuple{}$.
\end{compactitem}

Attributes labelling the same node in $\T$ have equal values in the
represented relation. The shape of $\T$ provides a hierarchy of
attributes by which we group the tuples of the represented relation:
we group the tuples by the values of the attributes labelling the
root, factor out the common values, and then continue recursively on
each group using the attributes lower in the f-tree. Branching into
several subtrees denotes a product of f-representations over the
individual subtrees. Examples~\ref{ex:running} and
\ref{ex:running-continued} give six f-trees and f-representations over
them.

For a given f-tree $\T$ over a set ${\cal S}$ of attributes, not all
relations over ${\cal S}$ have an f-representation over $\T$. However,
if a relation admits an f-representation $\Phi$ over $\T$, then $\Phi$
is unique up to commutativity of union and product.

\begin{example}\label{ex:dependent}
  The relation $R = \{\btuple{1,1},\btuple{1,2},\btuple{2,2}\}$ over
  schema $\{A,B \}$ does not admit an f-representation over the forest
  of f-trees $\{A\}$ and $\{B\}$, since there are no sets of values
  $a$ and $b$ such that $R$ is represented by $(\bigcup_a
  \data{A}{a})\times (\bigcup_b \data{B}{b})$. Its f-representation
  over the f-tree with root $A$ and child $B$ is
  $\data{A}{1}\times(\data{B}{1}\cup\data{B}{2})
  \cup\data{A}{2}\times\data{B}{2}$.\punto
\end{example}

\nop{For example, if $\T$ consists of two disjoint trees $\T_1$ and
  $\T_2$, any f-representation over $\T$ is a product of two
  expressions over $\T_1$ and $\T_2$, and hence the represented
  relation is always a product of two relations of the corresponding
  schemas.}


{\noindent\bf F-trees of a query.} Given a query $Q =\pi_{\cal
P}\sigma_\phi(R_1\times\cdots\times R_n)$, we can derive the f-trees
that define factorisations of the query result $Q(\D)$ for any input
database $\D$. We consider f-trees where nodes are labelled by
equivalence classes of attributes in ${\cal P}$; the equivalence class
of an attribute $A$ is the set of $A$ and all attributes transitively
equal to $A$ in $\phi$.

In addition, the attributes labelling the nodes have to satisfy a
so-called {\em path constraint}: all dependent attributes can only
label nodes along a same root-to-leaf path. The attributes of a
relation are dependent, since in general we cannot make any
independence assumption about the structure of a relation, cf.\@
Example~\ref{ex:dependent}. \nop{For instance, consider that two
attributes $A_1$ and $A_2$ of a relation label sibling nodes in an
f-tree $\T$ of $Q$ and let $n$ be their parent node. By definition,
for any value of $n$'s attributes, there must be several $A_1$-values
paired with several $A_2$-values, i.e., $A_1$ and $A_2$ are
independent conditioned on $n$. This independence assumption does not
hold in general; $A_1$ and $A_2$ are dependent and thus cannot label
sibling nodes in $\T$.} Attributes from different relations can also
be dependent. If we join two relations, then their non-join attributes
are independent conditioned on the join attributes. If these join
attributes are not in the projection list ${\cal P}$, then the
non-join attributes of these relations become dependent.

The path constraint is key to defining which f-trees represent valid
nesting structures for factorised query results.

\begin{proposition}\label{prop:valid-f-tree}
Given a query $Q$, an f-tree $\T$ of $Q$ satisfies the path constraint
if and only if for any input database $\D$ the query result $Q(\D)$
has an f-representation over $\T$.
\end{proposition} 

{\noindent\bf Tight size bounds for f-representations over f-trees.}
Given any f-tree $\T$, we can derive tight bounds on the size of
f-representa\-tions over $\T$ in polynomial time.

For any root-to-leaf path $p$ in $\T$, consider the hypergraph whose
nodes are the attributes classes of nodes in $p$ and whose edges are
the relations containing these attributes. The edge cover number of
$p$ is the minimum number of edges necessary to cover all attributes
in $p$. We can lift edge covers to their fractional
version~\cite{GM06}. The fractional edge cover number is the cost of
an optimal solution to the following linear program with variables
$\{x_{R_i}\}_{i=1}^n$:
\begin{align*}
\textrm{minimise}\quad&\textstyle\sum_i x_{R_i} \\
\textrm{subject to}\quad&\textstyle\sum_{i : R_i \in \rel({\cal A})} x_{R_i} \geq 1 \hspace*{.5em}\textrm{for all attribute classes ${\cal A}$,}\\
& x_{R_i} \geq 0 \hspace*{6em}\textrm{for all $i$.}
\end{align*} 

For each relation $R_i$ with attributes on $p$, its weight is given by
the variable $x_{R_i}$. Each attribute class ${\cal A}$ on $p$ has to
be covered by relations $\rel({\cal A})$ with attributes in ${\cal A}$
such that the sum of the weights of these relations is greater than 1.
The objective is to minimise the sum of the weights of all relations.
In the non-weighted version, the variables $x_{R_i}$ can only be
assigned the values 0 and 1, whereas in the weighted version, the
variables can be any positive rational number.

For an f-tree $\T$, we define $s(\T)$ as the maximum such fractional
edge cover number of any root-to-leaf path in $\T$.

\begin{example}
  Each f-tree $\T$ except for $\T_3$ in Figure~\ref{fig:ex-ftrees} has
  $s(\T) = 2$, while $s(\T_3) = 1$. In $\T_3$, both root-to-leaf paths
  $\textrm{supplier}-\textrm{item}$ and
  $\textrm{supplier}-\textrm{location}$ can be covered by relations
  $\textrm{Produce}$ and $\textrm{Serve}$ respectively.\punto
\end{example}

For any database $\D$ and f-tree $\T$, the size of the
f-represen\-tation of the query result over $\T$ is $O(|\D|^{s(\T)})$,
and there exist arbitrarily large databases $\D$ for which the size of
the f-representation over $\T$ is $\Omega(|\D|^{s(\T)})$. Given $\D$
and $\T$, f-representations of the query result $Q(\D)$ over the
f-tree $\T$ can be computed in time $O(|Q|\cdot|\D|^{s(\hat{\T})})$,
where $\hat{\T}$ is an extension of $\T$ with nodes for all attributes
in the input schema and not in the projection list ${\cal P}$;
detailed treatment of this result is given in prior
work~\cite{OZ12}. More succinct f-representations thus have a smaller
parameter $s(\T)$, which can be obtained by decreasing the length of
root-to-leaf paths in $\T$ and increasing the width of $\T$ while
preserving the path constraint.

We next define $s(Q)$ as the minimal $s(\T)$ for any f-tree $\T$ of
$Q$. Then, for any database $\D$, there is an f-representation of
$Q(\D)$ with size at most $|\D|^{s(Q)}$, and this is asymptotically
the best upper bound for f-representations over f-trees.

\begin{example}
  In Example~\ref{ex:running}, we have $s(Q_1) = 2$ since $Q_1$ admits
  no f-tree with $s(\T) < s(\T_1) = 2$. However, $s(Q_2) = 1$, since
  $\T_3$ is an f-tree of $Q_2$ and $s(\T_3) = 1$.\punto
\end{example}

The size bound $|\D|^{s(Q)}$ can be asymptotically smaller than the
size of the query result $Q(\D)$. For such queries, computing and
representing their result in factorised form can bring exponential
time and space savings in comparison to the traditional representation
as a set of tuples.

\begin{example}
  Consider relations $R_i$ over schemas $(A_i,B_i)$ and the query $Q_n
  = \sigma_{\Phi} (R_1\times\dots\times R_n)$, where $\Phi =
  \bigwedge_i (B_i = A_{i+1})$. This is a chain of $n-1$ equality
  joins. The result $Q_n(\D)$ can be as large as $|\D|^{\Theta(n)}$,
  while $s(Q_n) = \Theta(\log{n})$ and hence there exist factorised
  representations of $Q_n(\D)$ with size at most
  $|\D|^{\Theta(\log{n})}$. The value $s(\T) = \Theta(\log{n})$ is
  witnessed by an f-tree $\T$ with depth $\log{n}$.\punto
\end{example}


\begin{figure*}
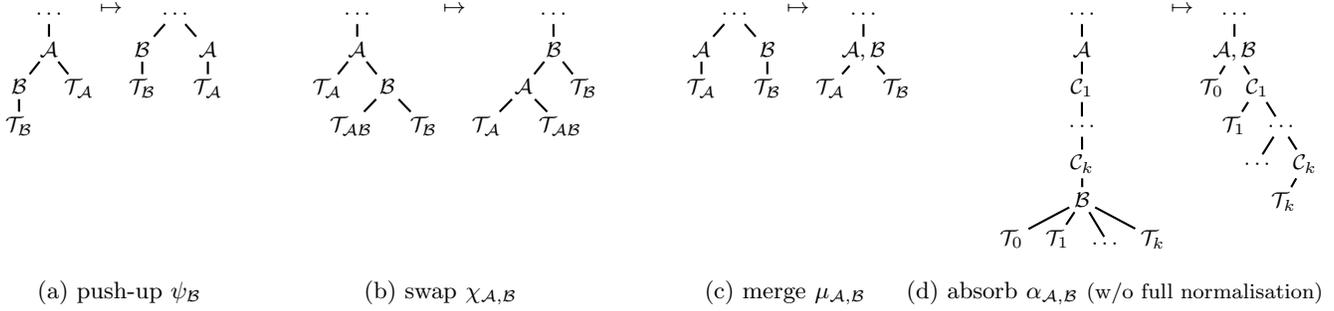

\begin{small}
\[
\psset{levelsep=5mm, nodesep=2pt, treesep=5mm}
\pstree{\TR{\cdots}}
{
  \pstree{\TR{\A}}
  {
    \pstree{\TR{\B}}
    {
      \TR{\T_\B}
    }
    \TR{\T_\A}
  }
}
\mapsto
\psset{levelsep=5mm, nodesep=2pt, treesep=5mm}
\pstree{\TR{\cdots}}
{
  \pstree{\TR{\B}}
  {
    \TR{\T_\B}
  }
  \pstree{\TR{\A}}
  {
    \TR{\T_\A}
  }
}
\qquad\qquad%
\psset{levelsep=5mm, nodesep=2pt, treesep=5mm}
\pstree{\TR{\cdots}}
{
  \pstree{\TR{\A}}
  {
    \TR{\T_\A}
    \pstree{\TR{\B}}
    {
      \TR{\T_{\A\B}}
      \TR{\T_\B}
    }
  }
}
\mapsto
\pstree{\TR{\cdots}}
{
  \pstree{\TR{\B}}
  {
    \pstree{\TR{\A}}
    {
      \TR{\T_\A}
      \TR{\T_{\A\B}}
    }
    \TR{\T_\B}
  }
}
\qquad\qquad%
\psset{levelsep=5mm, nodesep=2pt, treesep=5mm}
\pstree{\TR{\cdots}}
{
  \pstree{\TR{\A}}
  {
    \TR{\T_\A}
  }
  \pstree{\TR{\B}}
  {
    \TR{\T_\B}
  }
}
\mapsto
\psset{levelsep=5mm, nodesep=2pt, treesep=5mm}
\pstree{\TR{\cdots}}
{
  \pstree{\TR{\A,\B}}
  {
    \TR{\T_\A}
    \TR{\T_\B}
  }
}
\qquad\qquad%
\psset{levelsep=5mm, nodesep=2pt, treesep=3mm}
\pstree{\TR{\cdots}}
{
  \pstree{\TR{\A}}
  {
    \pstree{\TR{\C_1}}
    {
      \pstree{\TR{\cdots}}
      {
        \pstree{\TR{\C_k}}
        {
          \pstree{\TR{\B}}
          {
	    \TR{\T_0}
            \TR{\T_1}
            \TR{\dots}
            \TR{\T_k}
          }
        }
      }
    }
  }
}
\mapsto
\psset{levelsep=5mm, nodesep=2pt, treesep=3mm}
\pstree{\TR{\cdots}}
{
  \pstree{\TR{\A,\B}}
  {
    \TR{\T_0}
    \pstree{\TR{\C_1}}
    {
      \TR{\T_1}
      \pstree{\TR{\cdots}}
      {
        \TR{\cdots}
        \pstree{\TR{\C_k}}
        {
          \TR{\T_k}
          \phantom{\TR{X}}
        }
      }
    }
  }
}
\]
\end{small}

\hspace*{1em}
(a) push-up $\psi_\B$ 
\hspace*{6em}
(b) swap $\chi_{\A,\B}$
\hspace*{7em}
(c) merge $\mu_{\A,\B}$
\hspace*{1em}
(d) absorb $\alpha_{\A,\B}$ \begin{small}(w/o full normalisation)\end{small}

\caption{Transformations performed by f-plan operators depicted on f-trees.}
\label{fig:operators}\vspace*{-1em}
\end{figure*}

\section{Query Evaluation}
\label{sec:evaluation}

In this section we present a query evaluation technique on
f-representations. We propose a set of operators that map between
f-representations over f-trees. In addition to the standard relational
operators select, project, and Cartesian product, we introduce new
operators that can restructure f-representati\-ons and f-trees.
Restructuring is sometimes needed before selections, as exemplified in
the introduction. Any select-project-join query can be evaluated by a
sequential composition of operators called an f-plan.

We consider f-representations over f-trees as defined in
Section~\ref{sec:frep}. F-trees conveniently represent the structure
of factorisations as well as attributes and equality conditions on the
attributes. An f-tree uniquely determines (up to commutativity of
$\cup$ and $\times$) the f-representation of a given
relation. Therefore, the semantics of each of our operators may be
described solely by the transformation of f-trees $\T \mapsto \T'$. We
also present efficient algorithms to carry out the transformations on
f-representations. These algorithms are almost optimal in the sense
that they need at most quasilinear time in the sizes of both input and
output f-representations.


\begin{proposition}
The time complexity of each f-plan operator is $O(|\T|^2N\log N)$,
where $N$ is the sum of sizes of the input and output
f-representations and $\T$ is the input f-tree.
\end{proposition}

We assume that for any union expression $\bigcup_a$ in the input
f-representation, the values $a$ occur in increasing order, and that
the path constraint holds for the input f-tree.  Our algorithms
preserve these two constraints. 

We also introduce the notion of normalised f-trees, whose
f-representations cannot be further compacted by factoring out
subexpressions. We define an operator for normalising f-trees, and all
other operators expect normalised input f-trees and preserve
normalisation.

\subsection{Restructuring Operators}

{\noindent\bf The Normalisation Operator} factors out expressions
co\-mmon to all terms of a union. We first present a simple one-step
normalisation captured by the push-up operator $\psi_\B$, and then
normalise an f-tree by repeatedly applying the push-up operator
bottom-up to each node in the f-tree.

Consider an f-tree $\T$, a node $\A$ and its child $\B$ in $\T$. If
$\A$ is not dependent on $\B$ nor on its descendants, the subtree
rooted at $\B$ can be brought one level up (so that $\B$ becomes
sibling of $\A$) without violating the path
constraint. Proposition~\ref{prop:valid-f-tree} guarantees that there
is an f-representation over the new f-tree. Lifting up a node can only
reduce the length of root-to-leaf paths in $\T$ and thus decrease the
parameter $s(\T)$ and the size of the f-representation, cf.\@
Section~\ref{sec:frep}. Since the transformation only alters the
structure of the factorisation, the represented relation remains
unchanged.

Figure~\ref{fig:operators}(a) shows the transformation of
the relevant fragment of $\T$, where $\T_\A$ and $\T_\B$ denote the
subtrees under $\A$ and $\B$. F-representations over this fragment
have the form 
\[\Phi_1 = \textstyle\bigcup_a\; \big( \data{\A}{a} \times (\bigcup_b\; \data{\B}{b} \times F_b) \times E_a \big)\]
and change into
\[\Phi_2 = \textstyle(\bigcup_b\; \data{\B}{b} \times F_b) \times (\bigcup_a\; \data{\A}{a} \times E_a),\]
where each $E_a$ is over $\T_\A$, each $F_b$ is over $\T_\B$, and
$\data{\A}{a}$ stands for
$\data{A_1}{a}\times\cdots\times\data{A_n}{a}$ in case $A_1$ to $A_k$
are the attributes labelling node $\A$; the case of $\data{\B}{b}$ is
similar. Since neither $\B$ nor any node in $\T_\B$ depend on $\A$,
all copies of $(\bigcup_b\;\data{\B}{b} \times F_b)$ in $\Phi_1$ are
equal, so the transformation amounts to factoring out subexpressions
over the subtree rooted at $\B$. In any f-representation over $\T$,
the change shown above occurs for all unions over $\A$, and can be
executed in linear time in one pass over the f-representation.

\begin{definition}
An f-tree $\T$ is \emph{normalised} if no node in $\T$ can be pushed
up without violating the path constraint.
\end{definition}

Any f-tree $\T$ can be turned into a normalised one as follows. We
traverse $\T$ bottom up and push each node $\B$ and its subtree
upwards as far as possible using the operator $\eta_\B$. In case a
node $\A$ is pushed up, we mark it so that we do not consider it
again. If it is marked, so are all the nodes in its subtree, and at
least one of them is dependent on the parent of $\A$ (or $\A$ is a
root). The parent of $\A$ and its subtree do not change anymore after
$\A$ is marked, so $\A$ cannot be brought upwards again. All nodes are
marked after at most $|\T|^2$ applications of the push-up operator, so
the resulting f-tree is normalised. Since the size of the
f-representation over $\T$ decreases with each push-up, the time
complexity of normalising an f-representation is linear in the size of
the input f-representation. This procedure defines the normalisation
operator $\eta$.  In the remainder we only consider normalised f-trees
and operators that preserve normalisation.

\begin{example}
Let us normalise the left f-tree below with relations over schemas
$\{A,B\},\{B',C\},\{C',D\},$ $\{D',E\}$.
\begin{small}
\[
\psset{levelsep=6mm, nodesep=2pt, treesep=4mm}
\pstree{\TR{B,B'}}
{
  \pstree{\TR{A}}
  {
    \pstree{\TR{D,D'}}
    {
      \pstree{\TR{C,C'}}
      {
        \TR{E}
      }
    }
  }
}
\quad\mapsto\quad
\psset{levelsep=6mm, nodesep=2pt, treesep=4mm}
\pstree{\TR{B,B'}}
{
  \pstree{\TR{A}}
  {
    \pstree{\TR{D,D'}}
    {
      \TR{E}
      \TR{C,C'}
    }
  }
}
\quad\mapsto\quad
\psset{levelsep=6mm, nodesep=2pt, treesep=4mm}
\pstree{\TR{B,B'}}
{
  \pstree{\TR{D,D'}}
  {
    \TR{E}
    \TR{C,C'}
  }
  \TR{A}
}
\]
\end{small}
\vspace*{-1em}

The above transformation is obtained by $\psi_E$ followed by
$\psi_{\{D,D'\}}$. We can bring up $E$ since it is not dependent on
its parent in the left f-tree. We then mark $E$. We also mark
$\{C,C'\}$, since it cannot be brought upwards. The lowest unmarked
node is now $\{D,D'\}$. It can be brought upwards next to its parent
$A$ since $A$ is not dependent on it nor on any of its descendants.
The resulting f-tree is normalised.\punto
\end{example}

{\noindent\bf The Swap Operator} $\chi_{\A,\B}$ exchanges a node $\B$
with its parent node $\A$ in $\T$ while preserving the path constraint
and normalisation of $\T$. We promote $\B$ to be the parent of $\A$,
and also move up its children that do not depend on $\A$. The effect
of the swapping operator $\chi_{A,B}$ on the relevant fragment of $\T$
is shown in Figure~\ref{fig:operators}(b), where $\T_\B$ and
$\T_{\A\B}$ denote the collections of children of $\B$ that do not
depend, and respectively depend, on $\A$, and $\T_A$ denotes the
subtree under $\A$. Separate treatment of the subtrees $\T_\B$ and
$\T_{\A\B}$ is required so as to preserve the path constraint and
normalisation. The resulting f-tree has the same nodes as $\T$ and the
represented relation remains unchanged.

Any f-representation over the relevant part of the input f-tree $\T$
in Figure~\ref{fig:operators}(b) has the form
\[\textstyle\bigcup_{a} \left( \data{\A}{a} \times E_a \times \bigcup_{b} \left(\data{B}{b} \times F_b \times G_{ab} \right)\right),\]
while the corresponding restructured f-representation is
\[\textstyle\bigcup_{b} \left( \data{\B}{b} \times  F_{b} \times \bigcup_{a} \left(\data{\A}{a} \times E_{a} \times G_{ab} \right)\right).\]

The expressions $E_a$, $F_b$ and $G_{ab}$ denote the f-representati\-ons
over the subtrees $\T_\A$, $\T_\B$ and respectively $\T_{\A\B}$. 

The swap operator $\chi_{\A,\B}$ thus takes an f-representation where
data is grouped first by $\A$ then $\B$, and produces an
f-representation grouped by $\B$ then $\A$.
Figure~\ref{fig:algo-swap} gives an algorithm for $\chi_{\A,\B}$ that
executes this regrouping efficiently. We use a priority queue $Q$ to
keep for each value $a$ of attributes in $\A$ the minimal values $b$
of attributes in $\B$. This minimal value occurs first in the union
$U_a$ due to the order constraint of f-representations. We then
extract the values $b$ from the priority queue $Q$ in increasing order
to construct the union over them, and for each of them we obtain the
pairing values $a$. When a value $a$ is removed from $Q$, we insert it
back into $Q$ with the next value $b$ in its union $U_a$.

\nop{
We process each union over $\A$ separately. For each value $a$, we
maintain an iterator over the values $b$ that are in increasing order
and are multiplied with $\data{\A}{a}$ in the union $U_a$; we denote
by $p_a$ the current value $b$. We find the smallest $p_a$, call it
$b_{min}$, and all $a$'s for which $p_a = b_{min}$ in $Q$. We put
these $a$ values into a union multiplied by $\data{\B}{b}$. We then
proceed to the next element in the union multiplied by $\data{\A}{a}$,
and update $p(a)$. By repeating this process, we regroup the whole
fragment by $\B$.  }

Except for the operations on the priority queue, the total time taken
by the algorithm in any given iteration of the outermost loop is
linear in the size of the input $S_{in}$ plus the size of the output
$S_{out}$. For each $a$ in $S_{in}$ and $b$ in $U_a$, the value $a$ is
inserted into the queue with key $b$ once and removed once. There are
at most $|S_{in}|$ such pairs $(a,b)$ and each of the priority queue
operations runs in time $O(\log|S_{in}|)$. \nop{The time complexity of
the swapping operator is thus $O(|\mathrm{Input}|\log|\mathrm{Input}|
+ |\mathrm{Output}|)$.}

\begin{example}
\label{ex:swap}
The tree $\T_1$ in Figure~\ref{fig:ex-ftrees} is transformed into
$\T_2$ by the operator $\chi_{\textrm{item},\textrm{location}}$. \nop{The
subtree $\T_{\textrm{item},\textrm{location}}$ is empty.} The effect of
the operator on the f-representation amounts to regrouping it
primarily by $\textrm{location}$ instead of $\textrm{item}$, as
illustrated in Example~\ref{ex:running}.\punto
\end{example}

\begin{figure}[t]
\medskip\noindent
\framebox[\columnwidth]{
\parbox{8cm}{
\begin{small}
\mbox{\bf foreach} expression $S_{in}$ over the part of $\T$ in Figure~\ref{fig:operators}(b) {\bf do}\\
\hspace*{1em} create a new union $S_{out}$ \\
\hspace*{1em} let $Q$ be a min-priority-queue \\
\hspace*{1em} \mbox{\bf foreach} $\textstyle \data{\A}{a} \times E_a \times \bigcup_{b} \left(\data{\B}{b} \times F_b \times G_{ab} \right)$ in $S_{in}$  {\bf do}\\
\hspace*{2em} let $U_a$ be the union $\bigcup_{b} \left(\data{\B}{b} \times F_b \times G_{ab} \right)$ \\
\hspace*{2em} let $p_a$ be the first value $b$ in the union $U_a$ \\
\hspace*{2em} insert value $a$ with key $p_a$ into $Q$ \\
\hspace*{1em} \mbox{\bf while} $Q$ is not empty  {\bf do}\\
\hspace*{2em} let $b_{min}$ be the minimum key in $Q$ \\
\hspace*{2em} create a new union $V_{b_{min}}$ \\
\hspace*{2em} \mbox{\bf foreach} $a$ in $Q$ with key $b_{min}$  {\bf do}\\
\hspace*{3em} append $\data{\A}{a}\times E_a \times G_{ab}$ to $V_{b_{min}}$ \\
\hspace*{3em} remove $a$ from $Q$\\
\hspace*{3em} \mbox{\bf if} $p_a$ is not the last value in $U_a$ \mbox{\bf then} \\ 
\hspace*{4em} update $p_a$ to be the next value $b$ in the union $U_a$\\ 
\hspace*{4em} insert value $a$ with key $p_a$ into $Q$ \\
\hspace*{2em} append $\data{\B}{b_{min}} \times F_{b_{min}} \times V_{b_{min}}$ to $S_{out}$ \\
\hspace*{1em} replace $S_{in}$ by $S_{out}$
\end{small}
}
}
\caption{Algorithm for the swap operator $\chi_{\A,\B}$.}
\label{fig:algo-swap}
\vspace*{-1em}
\end{figure}

\subsection{Cartesian Product Operator}

Given two f-representations $E_1$ and $E_2$ over disjoint sets of
attributes, the product operator $\times$ yields the f-representation
$E = E_1 \times E_2$ over the union of the sets of attributes of $E_1$
and $E_2$ in time linear in the sum of the sizes of $E_1$ and $E_2$.
If $\T_1$ and $\T_2$ are the input f-trees, then the resulting f-tree
is the forest of $\T_1$ and $\T_2$.  It is easy to check that the
relation represented by $E$ is indeed the product of the relations of
$E_1$ and $E_2$, and that this operator preserves the constraints on
order of values, path constraint, and normalisation.

\subsection{Selection Operators}

We next present operators for selections with equality conditions of
the form $A = B$. Since equi-joins are equivalent to equality
selections on top of products, and the product of f-representations is
just their concatenation, we can evaluate equality joins in the same
way as equality conditions on attributes of the same relation, and do
not distinguish between these two cases in the sequel.

If both attributes $A$ and $B$ label the same node in $\T$, then by
construction of $\T$ the two attributes are in the same equivalence
class, and hence the condition $A=B$ already holds. If $\A$ and $\B$
are two distinct nodes labelled by $A$ and $B$ respectively in an
f-tree $\T$, the condition $A = B$ implies that $\A$ and $\B$ should
be merged into a single node labelled by the union of the  equivalence
classes of $A$ and $B$.

We propose two selection operators: the {\em merge} operator
$\mu_{\A,\B}$, which can only be applied in case $\A$ and $\B$ are
sibling nodes in $\T$, and the {\em absorb} operator $\alpha_{\A,\B}$,
which can only be applied in case $\A$ is an ancestor of $\B$ in
$\T$. For all other cases of $\A$ and $\B$ in $\T$, we first need to
apply the swap operator until we transform $\T$ in one of the above
two cases.  The reason for supporting these selection operators only
is that they are simple, atomic, can be implemented very efficiently,
and any selection can be expressed by a sequence of swaps and
selection operators. We next discuss them in depth.
\medskip

{\noindent\bf The Merge Selection Operator} $\mu_{\A,\B}$ merges the
sibling nodes $\A$ and $\B$ of $\T$ into one node labelled by the
attributes of $\A$ and $\B$ and whose children are those of $\A$ and
$\B$, see Figure~\ref{fig:operators}(c). This operator preserves the
path constraint, since the root-to-leaf paths in $\T$ are preserved in
the resulting f-tree.  Also, normalisation is preserved: merging two
nodes of a normalised f-tree produces a normalised f-tree. To preserve
the value order constraint, node merging is implemented as a
sort-merge join. Any f-representation over the relevant part of $\T$
has the form
\[\textstyle \Phi_1 = (\bigcup_a\; \data{\A}{a} \times E_a) \times (\bigcup_b\; \data{\B}{b} \times F_b),\]
and change into
\[\textstyle \Phi_2 = \bigcup_{a: a=b}\; \data{\A}{a} \times \data{\B}{b} \times E_a \times F_b,\]
where the union in $\Phi_2$ is over the equal values $a$ and $b$ of
the unions in $\Phi_1$.  An algorithm for $\mu_{\A,\B}$ needs one pass
over the input f-representation to identify expressions like $\Phi_1$,
and for each such expression it computes a standard sort-merge join on
the sorted lists of values of these unions.

\nop{
\medskip\noindent
\framebox[\columnwidth]{
\parbox{8cm}{
\begin{small}
\mbox{\bf foreach} subexpression $S_{in}$ over the shown fragment of $\T$ \\
\hspace*{2em} produce a new union $S_{out}$, \\
\hspace*{2em} \mbox{\bf repeat} \\
\hspace*{4em} find next $\data{\A}{a}$ and $\data{B}{b}$ in $S_{in}$ with $a=b$, \\
\hspace*{4em} append $\data{\A}{a} \times \data{\B}{b} \times E_a \times F_b$ to $S_{out}$, \\
\hspace*{2em} \mbox{\bf until} none $\data{\A}{a}$ or $\data{\B}{b}$ are left,\\
\hspace*{2em} replace $S_{in}$ by $S_{out}$.
\end{small}
}
}
}

\begin{example}
  Consider an f-tree that is the forest of $\T_1$ and $\T_4$ from
  Figure~\ref{fig:ex-ftrees}. The two attributes with the same name
  $\textrm{item}$ are siblings (at the topmost level). By merging
  them, we obtain the f-tree $\T_5$. Example~\ref{ex:running} shows
  f-representations over the input and output f-trees of this merge
  operation.\punto
\end{example}

{\noindent\bf The Absorb Selection Operator} $\alpha_{\A,\B}$ absorbs
a node $\B$ into its ancestor $\A$ in an f-tree $\T$, and then
normalises the resulting f-tree. The labels of $\B$ become now labels
of $\A$.

The absorption of $\B$ into $\A$ preserves the path constraint since
all attributes in $\B$ remain on the same root-to-leaf paths. By
definition, the absorb operator finishes with a normalisation step,
thus it preserves the normalisation constraint. Similar to the merge
selection operator, it employs sort-merge join on the values of $\A$
and $\B$ and hence creates f-representations that satisfy the order
constraint.

In any f-representation, each union over $\B$ is inside a union over
its ancestor $\A$, and hence inside a product with a particular value
$a$ of $\A$. Enforcing the constraint $A=B$ amounts to restricting
each such union over $\B$ by $B = a$, by which it remains with only
one or zero subexpression. This can be executed in one pass over the
f-representation, and needs linear time in the input size. The
subsequent normalisation also takes linear time. Both the absorption
and the normalisation only decrease the size of the resulting
f-representation.

\nop{
\medskip\noindent
\framebox[\columnwidth]{
\parbox{8cm}{
\begin{small}
\mbox{\bf foreach} subexpression $\data{A}{a}\times E_a$ of any union \\
\hspace*{2em} \mbox{\bf foreach} union $U$ over $\mathcal{B}$ inside $E_a$ \\
\hspace*{4em} erase all subexpressions $\data{B}{b}$ of $U$ for which $b \neq a$, \\
normalise the f-representation.
\end{small}
}
}
}

For normalising the f-tree after merging $\B$ into $\A$, we can use
the normalisation operator $\eta$ as described above. However, if the
original tree was normalised, it is sufficient to push up the subtrees
of $\B$ as shown in Figure~\ref{fig:operators}(d), but we may also
need to push upwards some of the nodes $\C_1, \dots, \C_k$ on the path
between $\A$ and $\B$.

\begin{example}
\label{ex:eq-ancestor}
Consider the selection $A = C$ on the leftmost f-tree below with
relations over schemas $\{A,B\}$, $\{B',C\}$ and $\{C',D\}$. Since $A$
and $C$ correspond to ancestor and respectively descendant nodes, we
can use the absorb operator to enforce the selection. When absorbing
$\{C,C'\}$ into $A$ (middle f-tree), the nodes $\{B,B'\}$ and $D$
become independent and $D$ can be pushed upwards (right f-tree):
\[
\psset{levelsep=6mm, nodesep=2pt, treesep=5mm}
\pstree{\TR{A}}
{
  \pstree{\TR{B,B'}}
  {
    \pstree{\TR{C,C'}}
    {
      \TR{D}
    }
  }
}
\qquad\mapsto\qquad
\psset{levelsep=6mm, nodesep=2pt, treesep=5mm}
\pstree{\TR{A,C,C'}}
{
  \pstree{\TR{B,B'}}
  {
    \TR{D}
  }
}
\qquad\mapsto\qquad
\psset{levelsep=6mm, nodesep=2pt, treesep=5mm}
\pstree{\TR{A,C,C'}}
{
  \TR{B,B'}
  \TR{D}
}\vspace*{-1em}
\]\punto
\end{example}
\vspace*{-1em}

{\noindent\bf The Selection with Constant Operator} $\sigma_{A\theta
c}$ can be evaluated in one pass over the input f-representation
$E$. Whenever we encounter a union $\bigcup_a (\data{A}{a}\times E_a)$
in $E$, we remove all expressions $\data{A}{a}\times E_a$ for which $a
\neg\theta c$. If the union becomes empty and appears in a product
with another expression, we then remove that expression too and
continue until no more expressions can be removed. In case $\theta$ is
an equality comparison, then all remaining $A$-values are equal to $c$
and we can factor out the singleton $\data{A}{c}$.

For a comparison $\theta$ different from equality, the f-tree remains
unchanged. In case of equality, we can infer that all $A$-values in
the f-representation are equal to $c$ and thus the node $\A$ labelled
by $A$ is independent of the other nodes in the f-tree and can be
pushed up as the new root. When computing the parameter $s(\T)$, we
can ignore $\A$ since the only f-representation over it is the
singleton $\data{A}{c}$.
\medskip

\nop{
{\noindent\bf Theta conditions.} Currently, FDB can only consider
theta-joins via an indirection. We can evaluate an arbitrary condition
$A\theta B$ on a relation $R$ (which may be the result of subsequent
joins) by joining $R$ with a binary relation $\Theta_{A,B} =
\sigma_{A\theta B}(R)$ on the attributes $A$ and $B$. Any such join
can be incorporated in our system by introducing the extra relation
$\Theta_{A,B}$, which need not be materialised in the database, and
hence an extra dependency set $\{A,B\}$.
}

\subsection{Projection Operator}

Given an f-representation $E$, the projection operator
$\pi_{\bar{A}}$, where $\bar{A}$ is a list of attributes of $E$,
replaces singletons $\data{B}{b}$ of type $B\not\in\bar{A}$ with the
empty singleton $\langle\rangle$. If an empty singleton appears in a
product with other singletons, then it can be removed from $E$. Also,
a union of empty singletons is replaced by one empty singleton. This
procedure can be performed in one scan over the input f-representation
$E$ and trivially preserves the order constraint.

We transform the input f-tree as follows. We first mark those
attributes that are projected away without removing them from the
f-tree. The set of attributes of an f-tree would then exclude the
marked attributes. If a leaf node has all attributes marked, we may
then remove the node and its attributes from the f-tree. This process
is repeated until no more nodes can be removed. We do not remove inner
nodes with all attributes marked for the following reason. Consider
the f-tree $\T$ representing a path $A-B-C$ and with dependency sets
$\{A,B\}$ and $\{B,C\}$. Now assume that we project away the attribute
$B$. If we would completely remove $B$ from $\T$, the nodes $A$ and
$C$ would become independent in the resulting f-tree, and we could
then normalise it into a forest of nodes $A$ and $C$. However, this is
not correct. The nodes $A$ and $C$ still remain {\em transitively}
dependent on each other. We therefore swap nodes such that those with
all attributes marked become leaves, in which case we can remove them
as explained above. The projection operator trivially preserves the
path constraint and normalisation.


\section{Query Optimisation}
\label{sec:optimisation}

In this section, we discuss the problem of query optimisation for
queries on f-representations. In addition to the optimisation
objective present in the standard (flat) relational case, namely
finding a query plan with minimal cost, the nature of factorised data
calls for a new objective: from the space of equivalent
f-representations for the query result, we would like to find a small,
ideally minimal, f-representation.

The operators described in Section~\ref{sec:evaluation} can be
composed to define more complex transformations of f-representations
over f-trees.  Any select-project-join query can be evaluated by
executing a sequence of these operators. Such a sequence of operators
is called an f-plan and several f-plans may exist for a given
query. In this section we introduce different cost measures for
f-plans and algorithms for finding optimal ones.

The products and selections with constant are the cheapest on
f-representations and can be evaluated first using the corresponding
operators. Projection can only be evaluated when the nodes with no
projection attributes are leaves of the f-tree, and in FDB they are
deferred until the end. Most expensive are the equality selection
operators and the restructuring operators which make selections and
projections possible. Their evaluation order is addressed in this
section.

A selection ${A=B}$ can only be executed on an f-representa\-tion over
an f-tree $\T$ if the attributes $A$ and $B$ label nodes $\A$ and
respectively $\B$ that are either the same, siblings, or along a same
path in $\T$. Otherwise, we first need to transform the
f-representation. If $\A$ and $\B$ are in the same tree, we can e.g.\
repeatedly swap $\A$ with its parent until it becomes an ancestor of
$\B$. If $\A$ and $\B$ are in disjoint trees of $\T$ (recall that $\T$
may be a forest), we can promote both of them as roots of their
respective trees by repeatedly swapping nodes, and thus as siblings at
the topmost level in the f-tree. To complete the evaluation, we apply
a merge or absorb selection operator on the two nodes $\A$ and $\B$.

There are several choices involved in the evaluation of a conjunction
of selection conditions: For each selection, should we transform the
input f-tree, and consequently the f-representation, such that the
nodes $\A$ and $\B$ become siblings or one the ancestor of the other?
Is it better to push up $\A$ or $\B$? What is the effect of a
transformation for one selection on the remaining selections? The aim
of FDB's optimiser is to find an f-plan for the given query such that
the maximal cost of the sequence of transformations is low and the
query result is well-factorised.

\subsection{Cost of an F-Plan}
\label{sec:cost}

We next define two cost measures for f-plans. One measure is based on
the parameter $s(\T)$ that defines size bounds on factorisations over
f-trees for any input database. The second measure is based on
cardinality estimates inferred from the intermediate f-trees
\emph{and} catalogue information about the database, such as relation
sizes and selectivity estimates. Both measures can be used by the
exhaustive search procedure and the greedy heuristic for query
optimisation presented later in this section.

\nop{\new{
Each of the operators presented Section~\ref{sec:evaluation} has
complexity linear in the input and output f-representation sizes,
except for the restructuring operators, whose complexity is
quasilinear. Our cost measures for f-plans are therefore based on cost
measures for the intermediate f-representations produced by the
individual f-plan operators.  }}

{\noindent\bf Cost Based on Asymptotic Bounds.} As discussed in
Section~\ref{sec:frep}, the size of any f-representation over an
f-tree $\T$ depends exponentially on the parameter $s(\T)$, i.e., the
size is in $O(|\D|^{s(\T)})$. Since the cost of each operator is
quasilinear in the sum of sizes of its input and output, the parameter
$s(\T)$ dictates it. For an f-plan $f$ consisting of operations
$\omega_1,\dots,\omega_k$ that transform f-representations and their
f-trees:
\[\T_{\mathrm{initial}} = \T_0 \:\:{\buildrel \omega_1 \over \mapsto}\:\: \T_1 \:\:{\buildrel \omega_2 \over \mapsto}\:\: \dots \:\:{\buildrel \omega_k \over \mapsto}\:\: \T_k = \T_{\mathrm{final}},\]
the evaluation time is $O(|\D|^{s(f)}\cdot\log{|\D|})$, where
$$s(f) = \max(s(\T_0),s(\T_1),\dots,s(\T_k)).$$

The sizes of the intermediate f-representations thus dominate the
execution time. Using this cost measure, a good f-plan is one whose
intermediate f-trees $\T_i$ have small $s(\T_i)$.

In defining a notion of optimality for f-plans, we would like to
optimise for two objectives, namely minimise $s(f)$ and
$s(\T_{\mathrm{final}})$. However, it might not be possible to
optimise for both objectives $<_{\max}$ and $<_{s(\T)}$ at the same
time.  Instead, we set for an order on these objectives. We define
the lexicographic order $<_{\max}\times <_{s(\T)}$ on f-plans
consisting of the following orders:
\begin{compactenum}
\item $f_1 <_{\max} f_2$ holds if $s(f_1) < s(f_2)$, and
\item $f_1 <_{s(\T)} f_2$ holds if $s(\T_1)< s(\T_2)$, where $\T_1$ 
and $\T_2$ are the f-trees of the query result computed by $f_1$ and
$f_2$ respectively.
\end{compactenum}
Given f-plans $f_1$ and $f_2$, we consider $f_1$ better than $f_2$ and
write $f_1 <_{\max}\times <_{s(\T)} f_2$ if either (1) the most
expensive operator in $f_1$ is less expensive than the most expensive
operator in $f_2$, or (2) their most expensive operators have the same
cost but the cost of the result is smaller for $f_1$. An f-plan $f_1$
for a query $Q$ is {\em optimal} if there is no other f-plan $f_2$ for
$Q$ such that $f_2 <_{\max}\times<_{s(\T)} f_1$.

This notion of optimality is over f-plans consisting of operators defined in Section~\ref{sec:evaluation}. Since these operators preserve f-tree normalisation, this also means that we consider optimality only over the space of possible normalised
f-trees.

\begin{example}
\label{ex:fplan}
Consider the following f-plan evaluating the selection $B=F$ on the leftmost f-tree, with dependency sets $\{A,B,C\}$ and $\{D,E,F\}$.
\[
\psset{levelsep=6mm, nodesep=2pt, treesep=5mm}
\pstree{\TR{A,D}}
{
  \pstree{\TR{B}}
  {
    \TR{C}
  }
  \pstree{\TR{E}}
  {
    \TR{F}
  }
}
\qquad{\buildrel \chi_{\{A,D\},B} \over \mapsto}\qquad
\psset{levelsep=6mm, nodesep=2pt, treesep=5mm}
\pstree{\TR{B}}
{
  \pstree{\TR{A,D}}
  {
    \TR{C}
    \pstree{\TR{E}}
    {
      \TR{F}
    }
  }
}
\qquad{\buildrel \alpha_{B,F} \over \mapsto}\qquad
\psset{levelsep=6mm, nodesep=2pt, treesep=5mm}
\pstree{\TR{B,F}}
{
  \pstree{\TR{A,D}}
  {
    \TR{C}
    \TR{E}
  }
}
\]
The input f-tree and the output f-tree have both cost 1, as each
root-to-leaf path is covered by a single relation. However, the
intermediate f-tree has cost 2 (as on the path from $B$ to $F$ each of
$B$ and $F$ must be covered by a separate relation), so the cost of
the f-plan is 2. An alternative f-plan starts by swapping $F$ with its
parent to obtain an intermediate f-tree with cost 1, and then merges
$F$ with $B$.
\[
\psset{levelsep=6mm, nodesep=2pt, treesep=5mm}
\pstree{\TR{A,D}}
{
  \pstree{\TR{B}}
  {
    \TR{C}
  }
  \pstree{\TR{E}}
  {
    \TR{F}
  }
}
\qquad{\buildrel \chi_{E,F} \over \mapsto}\qquad
\psset{levelsep=6mm, nodesep=2pt, treesep=5mm}
\pstree{\TR{A,D}}
{
  \pstree{\TR{B}}
  {
    \TR{C}
  }
  \pstree{\TR{F}}
  {
    \TR{E}
  }
}
\qquad{\buildrel \mu_{B,F} \over \mapsto}\qquad
\psset{levelsep=6mm, nodesep=2pt, treesep=5mm}
\pstree{\TR{A,D}}
{
  \pstree{\TR{B,F}}
  {
    \TR{C}
    \TR{E}
  }
}
\]
Although both f-plans result in an f-tree with cost 1\nop{ (we can
convert between the resulting f-trees using
$\chi_{\{A,D\},\{B,F\}}$)}, the latter f-plan has cost 1 while the
former has cost 2.\punto
\end{example}

{\noindent\bf Cost Based on Estimates.} We can also
estimate the cost of an f-plan computing the factorised query result
for a query $Q$ and database $\D$ using cardinality and selectivity
estimates for $\D$.

Given an f-representation $E$ over an f-tree $\T$ of a query $Q$ and
an attribute $A$ in $\T$, the number of $A$-singletons in $E$ is given
by the size of the result of a query $Q_{\anc(A)}$ on the input
database $\D$. This query is $\pi_{\anc(A)}(Q)$, where $\anc(A)$ is
the set of attributes labelling nodes from the root to the node of $A$
in $\T$~\cite{OZ12}. For instance, in Example~\ref{ex:running}, the
number of occurrences of any {\tt dispatcher} in the first
f-representation over the f-tree $\T_1$ is the number of combinations
of values for {\tt item-location-dispatcher} in the query result.

The size of the factorisation $E$ is then $\sum_{A \in \mathcal{P}}
|Q_{\anc(A)}(\D)|$ over all attributes $A$ in the projection list
${\cal P}$ of $Q$. The cardinality of $Q_{\anc(A)}(\D)$ can now be
estimated using known techniques for relational databases,
e.g.,~\cite{Gehrke2003}. The cost $s(f)$ of an f-plan $f$ can be
estimated as the sum of the cost estimates of the intermediate and
final f-trees.

Given an f-tree $\T$ and database estimates, we need polynomial time
in $\T$ to find $s(\T)$ using linear programming and to compute the
size estimate.

\begin{figure*}[t]
  \centering
  \subfloat{
    \includegraphics{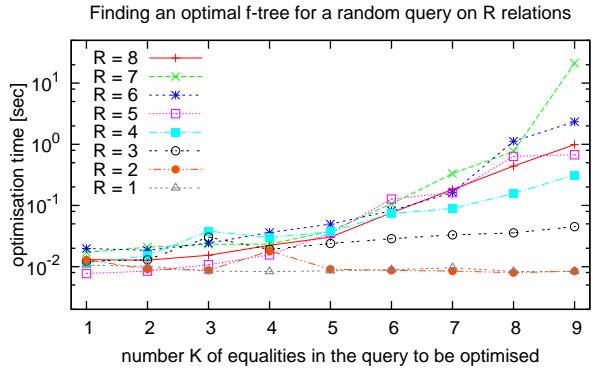}
  }
  \subfloat{
    \includegraphics{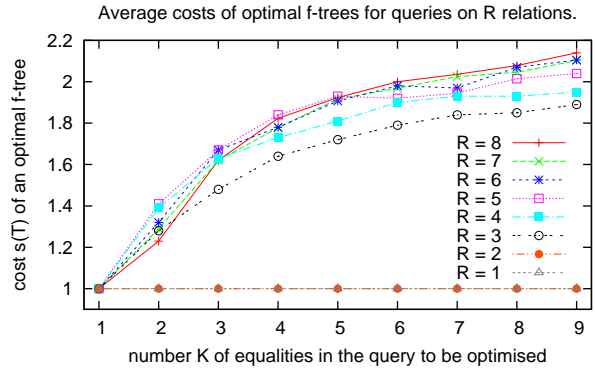}
  }
  \vspace*{-0.5em}
  \caption{Experiment 1: Query optimisation on flat data, $K$ equalities on $R$ relations with $A = 40$ attributes.}
  \label{fig:exp1}
  \vspace*{-1em}
\end{figure*}

\subsection{Exhaustive Search}

To find an optimal f-plan for an equi-join query we search the space
of all possible normalised f-trees and all possible operators between
the f-trees (thus represented as a directed graph where f-trees are
nodes and operators are edges). An f-plan for a given selection query
$Q$ on an input over an f-tree $\T_{in}$ is any path $f$ from
$\T_{in}$ to some final f-tree $\T_{final}$ such that (1) the
equivalence classes of $\T_{final}$ are the classes of $\T_{in}$ joined
by the query equalities.

The cost function $s(f)$ defines a distance function on the space of
f-trees: the distance from $\T_1$ to $\T_2$ is the minimum possible
cost $s(f)$ of an f-plan from $\T_1$ to $\T_2$. We are thus searching
for f-trees $\T_{final}$ which satisfy (1), are closest to $\T_{in}$
(2), and have smallest possible cost (3). We can use Dijkstra's
algorithm to find distances of all f-trees from $\T_{in}$: explore the
space starting with the $\T_{in}$ and trying all allowed operators,
processing the reached f-trees in the order of increasing distance
from $\T_{in}$. Then, among all f-trees satisfying (1), we pick one
with the shortest distance from $\T_{in}$, and among these we pick one
with smallest cost. Then we output a shortest path from $\T_{in}$ to
this f-tree.

\nop{\new{
In the presence of catalogue information and cardinality estimates
$e(f)$ for plans, the only change to the search algorithm is in the
metric. The cost $s(f)$ combines using the max-metric while the cost
$e(f)$ combines additively, but an adaptation of Dijkstra's algorithm
can be used to find the shortest path in either metric.  }}

The complexity of the search is determined by the size of the search
space. By successively applying operators to $\T_{in}$, we rearrange
its nodes (swap operator) or merge pairs of its nodes (merge and
absorb operators). For each partition of attributes over nodes, there
will be a cluster of f-trees with the same nodes but different shape,
among which we can move (transitively) using the swap operator. By
applying a merge or absorb operator, we move to a cluster whose
f-trees have one fewer node. Since we can never split a node in two,
any valid f-plan will only merge nodes which end up merged in
$\T_{final}$. For a query with $k$ equality selections, there are at
most ${k+1 \choose 2 }=O(k^2)$ pairs of nodes we may merge and we
perform at most $k$ merges, so there are $O(k^{2k})$ reachable
clusters. In a cluster with $m$ nodes there are at most $m^m$ f-trees.
Since $m$ will be always at most the size $n$ of the initial f-tree
$\T_{in}$, the size of the search space is $O(k^{2k}n^n)$.

\subsection{Greedy Heuristic}

Our greedy optimisation algorithm restricts the search space for
f-plans in two dimensions: (1) it only applies restructuring operators
to nodes that participate in selection conditions, and (2) it
considers a standard greedy approach to join ordering, whereby at each
step it chooses a join with the least cost from the remaining joins.

The algorithm constructs an f-plan $f$ for a conjunction of equality
conditions as follows. For each condition involving two attributes
labelling nodes $\A$ and $\B$, we consider three possible
restructuring scenarios: swapping one of the nodes $\A$ and $\B$ until
$\A$ becomes the ancestor of $\B$ or the other way around, or bringing
both $\A$ and $\B$ upwards until they become siblings. We choose the
cheapest f-plan for each condition. This f-plan involves restructuring
followed by a selection operator to perform the condition. We then
order the conditions by the cost of their f-plans. The condition with
the cheapest f-plan is performed first and its f-plan is appended to
the overall f-plan $f$. We then repeat this process with the remaining
conditions until we finish them. The new input f-tree is now the
resulting f-tree of the f-plan of the previously chosen condition.

In contrast to the full search algorithm, this greedy algorithm takes
only polynomial time in the size of the input f-tree $\T$. For each
condition, there can be at most $O(|\T|)$ swaps and each swap requires
to look at all descendants of the swapped nodes to check for
independence. Computing the resulting f-tree in each of the three
restructuring cases would then need $O(|\T|^2)$.


\section{Experimental Evaluation}
\label{sec:experiments}

\begin{figure*}[t]
  \includegraphics[scale=1.4]{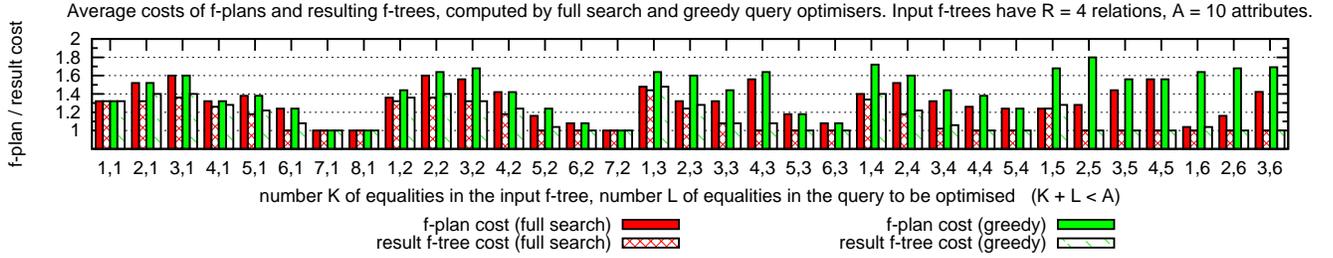}
  \vspace*{-2em}
  \caption{Experiment 2: Comparison of full-search and greedy query optimisers.}
  \label{fig:exp2-ss}
  \vspace*{-1em}
\end{figure*}

\begin{figure*}[t!]
  \center
  \subfloat{
    \includegraphics{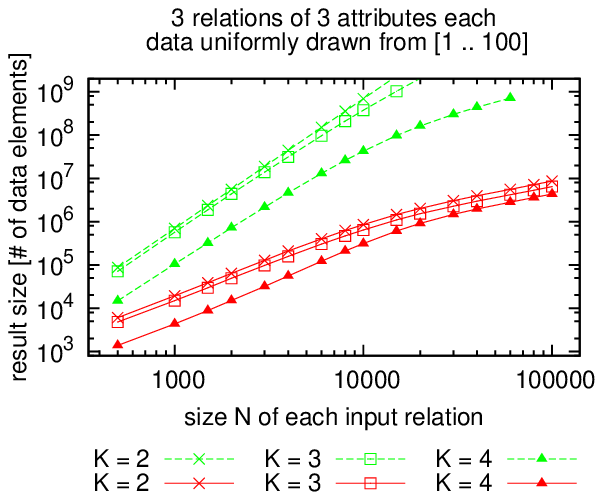}
  }\hspace*{-2.5em}
  \subfloat{
    \includegraphics{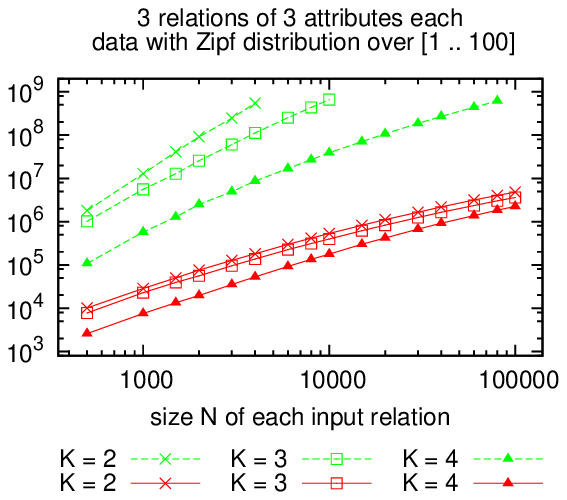}
  }\hspace*{-2em}
   \raisebox{2.8em}{
    \subfloat{
    \includegraphics{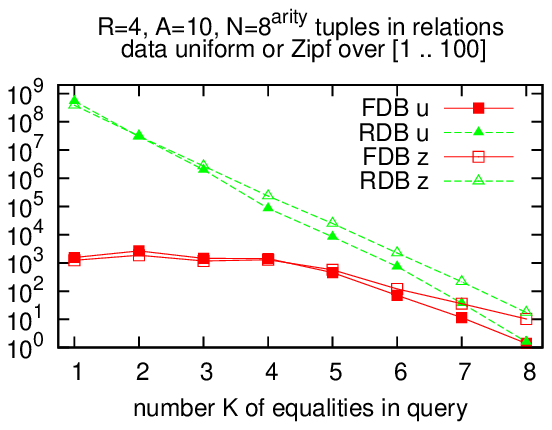}
  }}\vspace*{-1em}

  \subfloat{
    \includegraphics{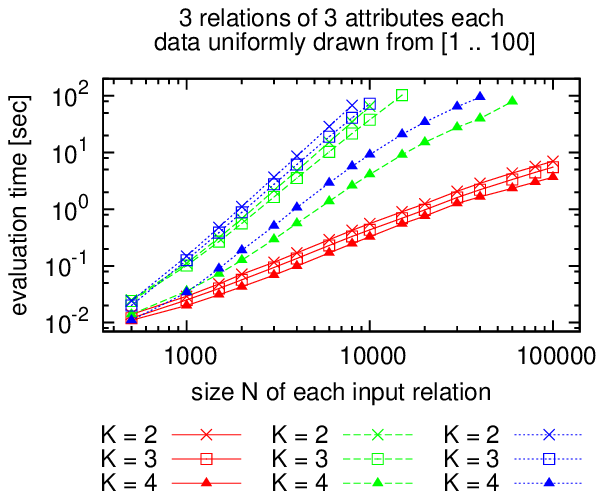}
  }\hspace*{-2.5em}
  \subfloat{
    \includegraphics{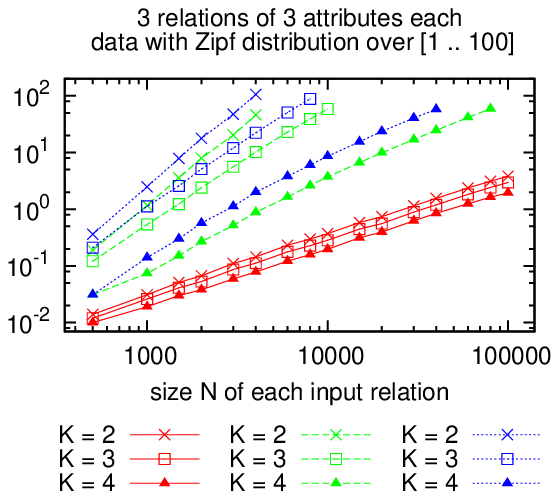}
  }\hspace*{-2em}
  \raisebox{2.9em}{\subfloat{
    \includegraphics{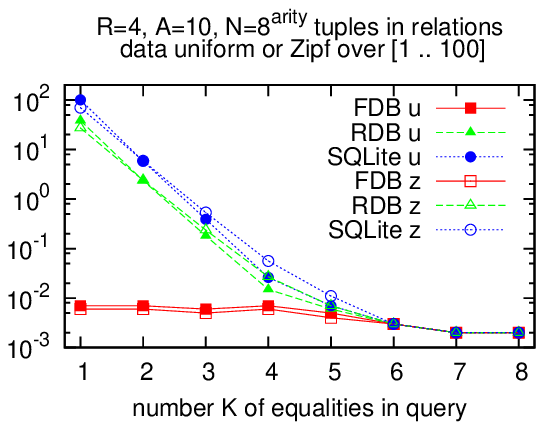}
  }}  \vspace*{-1em}

  \caption{Experiment 3: Performance of query evaluation on flat relational data. For sizes (top row): FDB (solid red), RDB and SQLite and PostgreSQL (dashed green). For times (bottom row): FDB (solid red), RDB (dashed green), and SQLite (dotted blue); PostgreSQL is ca.\@ 3 times slower than SQLite and not shown. }
  \label{fig:exp3}
  \vspace*{-1em}
\end{figure*}

\begin{figure*}[t!]
  \subfloat{
    \includegraphics{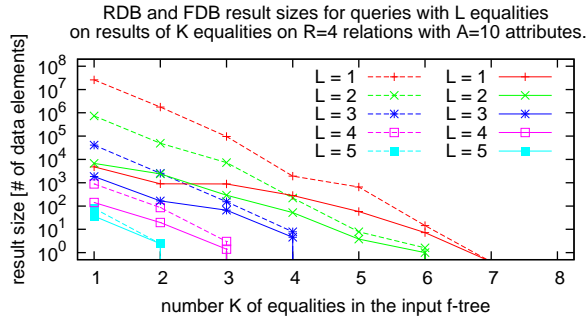}
  }
  \subfloat{
    \includegraphics{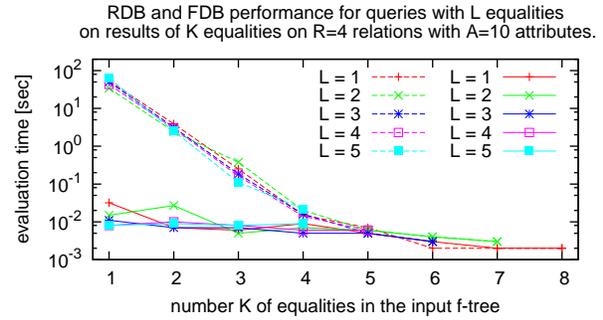}
  }
  \caption{Experiment 4: Performance of FDB (solid lines) and RDB
    (dashed lines) for query evaluation on factorised data. RDB needs
    one scan over the input relation, whereas FDB may need
    restructuring the factorised input. The times and result sizes for
    SQLite and PostgreSQL are as for RDB and not plotted.}
  \label{fig:exp4}
  \vspace*{-1em}
\end{figure*}

\begin{figure}[t]
  \includegraphics{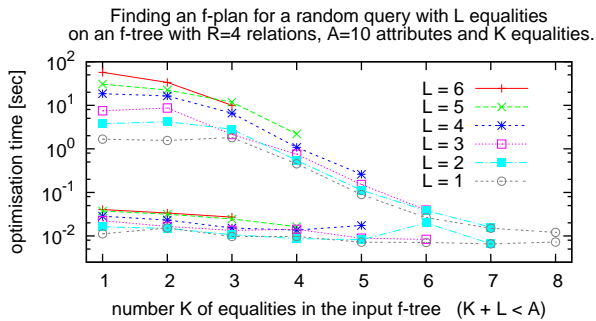}
  \caption{Experiment 2: Performance comparison of query optimisers.
    Slower series (top) correspond to full search, faster series
    (bottom) to greedy.}
  \label{fig:exp2-time}
  \vspace*{-1em}
\end{figure}

We evaluate the performance of our query engine FDB against three
relational engines: one homebred in-memory (RDB) and two open-source
engines (SQLite and Postgre\-SQL). Our main finding is that FDB
clearly outperforms relational engines for data sets with many-to-many
relationships. In particular, in our experiments we found that:
\begin{compactitem}
\item The size of factorised query results is typically at most
  quadra\-tic in the input size for queries of up to eight relations
  and nine join conditions (Figure~\ref{fig:exp1} right).

\item Finding optimal f-trees for queries of up to eight relations
  and six join conditions takes under 0.1 seconds (Figure~\ref{fig:exp1} left).
  Finding optimal f-plans for queries on factorised data is about an
  order of magnitude slower. In contrast, the greedy optimiser takes under 5 ms
  (Figure~\ref{fig:exp2-time}) without any significant loss in the
  quality of factorisation (Figure~\ref{fig:exp2-ss}).

\item For queries on input relations, factorised query results are two
  to six orders of magnitude smaller than their flat equivalents and
  FDB outperforms RDB by up to four orders of magnitude
  (Figure~\ref{fig:exp3}). \nop{\new{The gap increases with increasing
      data size.}}  For the same workload SQLite performed about three
  times slower than RDB, and PostgreSQL performed three times slower
  than SQLite; both systems have additional overhead of fully
  functioning engines. Also, RDB implements a hand-crafted optimised
  query plan.

\item The above observations hold for both uniform and Zipf data
  distributions, with a slightly larger gap in performance for the
  latter (Figure~\ref{fig:exp3}).

\item The evaluation of subsequent queries on input data representing
  query results has the same time performance gap, since the new input
  is more succinct as factorised representation than as relation.
  Figure~\ref{fig:exp4} compares evaluation times for selection
  queries on (1) one relation, which can be trivially evaluated by a
  single scan of this relation, and on (2) the factorisation of that
  relation, which may require restructuring.

\item  For one-to-many (e.g., key-foreign key) relationships, the 
  performance gap is smaller, since the result sizes for one-to-many
  joins can only depend linearly on the input size and not
  quadratically as in the case of many-to-many joins and the possible
  gain brought by factorisation is less dramatic. For instance, in the
  TPC-H benchmark, the joins are predominantly on keys and therefore
  the sizes of the join results do not exceed that of the relation
  with foreign keys. Factorised query results are still more succinct
  than their relational representations, but only by a factor that is
  approximately the the number of relations in the query (experiments
  not plotted due to lack of space).
\end{compactitem}

{\noindent\bf Competing Engines.} We implemented FDB and RDB in C++
for execution in main memory, using the GLPK package v4.45 for solving
linear programs. FDB evaluation and optimisation are described in
previous sections. We also used the lightweight query engine SQLite
3.6.22 tuned for main memory operation by turning off the journal mode
and synchronisations and by instructing it to use in-memory temporary
store. Similarly, we run PostgreSQL 9.1 with the following parameters:
fsync, synchronous commit, and full page writes are off, no background
writer, shared buffers and working memory increased to 12 GB. Both
SQLite and PostgreSQL read the data in their internal binary format,
whereas FDB and RDB use the plain text format. The relations are given
sorted; this allows RDB to use optimal relational join plans
implemented as multi-way sort-merge joins. For all engines we report
wall-clock times (averaged over five runs) to read data from disk and
execute the query plans without writing the result to disk.

{\noindent\bf Experimental Setup.} All experiments were performed on
an Intel(R) Xeon(R) X5650 Quad 2.67GHz/64bit/32GB running VMWare VM
with Linux 2.6.32/gcc4.4.5.

{\noindent\bf Experimental Design.}  The flow of our experiments is as
follows. We generate random data and queries, then repeat a number of
times four optimisation and evaluation experiments and report averages
of optimisation time, execution time, representation sizes, and
quality of produced f-plans.

We generate $R$ relations and distribute uniformly $A$ attributes over
them. Each relation has a given number of tuples, each value is a
natural number generated from 1 to $M$ using uniform or Zipf
distribution. The queries are equi-joins over all of these relations.
Their selections are conjunctions of $K$ non-redundant equalities.

For each generated query $Q$ and database $\D$, we do the following.
In Experiment 1, we run the FDB optimiser to find an optimal f-tree
$\T$ for the query result and report the optimisation time and the
value of the parameter $s(Q)$ that controls the size of the
f-representation of $Q(\D)$ over $\T$. In Experiment 3, we compute the
result $Q(\D)$ using RDB, SQLite, and PostgreSQL, and the factorised
query result using FDB. We then report on both the evaluation time and
size of the result as the number of its singletons; a singleton holds
an 8 byte integer.

In Experiments 2 and 4, we consider new queries on top of results
produced in Experiments 1 and 3 respectively. The new queries are also
equi-joins, where the selections are conjunctions of $L$ random (not
already implied) equalities on attribute equivalence classes of $\T$.

For each new query $Q'$, we run the FDB optimiser to find an optimal
f-plan to compute the result and the resulting f-tree of the query
result. In Experiment 2, we report the optimisation time and quality
of the computed f-plans with the exhaustive and greedy optimisation
algorithms; here, we consider the cost of the f-plan defined by the
parameter $s(\cdot)$ of the intermediate and final f-trees; in our
experiments, the alternative cost estimate discussed in
Section~\ref{sec:cost} would lead to very similar choices of optimal
f-plans. In Experiment 4, we execute the chosen f-plan with FDB and
compute with the relational engines the selection conditions given by
$Q'$ on a {\em single} relation $Q(\D)$ computed in Experiment 3. We
report the execution times and query result sizes.

The parameters $K$ and $L$ are subject to $K+L<A$, as we can do at
most $A-1$ non-trivial joins on $A$ attributes. We run the experiments
five times for each parameter setting.  

{\noindent\bf Experiment 1: Query optimisation on flat data.}

Figure~\ref{fig:exp1} shows average times for optimising a query on
flat data, and average costs $s(\T)$ for the chosen optimal f-tree
$\T$ of the query result. For schemas with $A=40$ attributes over
$R=1,\dots,8$ relations, we optimised queries of $K=1,\dots,9$
equality selections. The cost $s(\T)$ of an optimal f-tree $\T$ is
always $1$ for queries of up to two relations. For $R \geq 3$ and a
sufficient number of joins we often get queries with optimal $s(\T)=2$
and in very rare cases $s(\T) > 2$.  This means that the sizes of
f-representations for the query results are in most cases quadratic in
the size of the input database even in the case of 9 equality
selections on 8 relations. The optimiser searches a potentially
exponentially large space of f-trees to find an optimal one, but runs
under 1 second for queries with less than $8$ joins on up to $8$
relations.

{\noindent\bf Experiment 2: Query optimisation on factorised data.}

Figure~\ref{fig:exp2-ss} shows the behaviour of query optimisation for
factorised data. It shows the costs of the computed f-plans as well as
the costs of the f-tree of the result computed by the f-plans for our
full-search and greedy optimisation algorithms.

The queries under consideration have $L\leq 6$ joins on
f-representati\-ons resulting after $K\leq8$ joins on $R=4$ relations
with $A=10$ attributes. The greedy optimiser gives optimal or nearly
optimal results in most cases (by comparison with the optimal outcome
of full search). The exceptions are queries joining most attributes of
an f-representation produced by a query with few joins (small $K$,
large $L$). In all cases the average f-plan cost is between 1 and 2,
which means that the f-plans produce factorisations of at most
quadratic size even though we join 4 relations. The results also show
that for small queries (small $L$) the cost of the optimal f-plan is
dominated by the cost of the final f-tree. As the query size (i.e.,
$L$) grows, the result f-tree has less attribute classes and its cost
is smaller than the cost of the f-plan (i.e., smaller than the cost of
intermediate f-trees that we must process while evaluating the query).

Figure~\ref{fig:exp2-time} shows the execution times for both
optimisers. The search space of possible f-trees grows exponentially
with the number $L$ of selections and also with the size of the input
f-tree (i.e., with decreasing $K$).  The performance of the
full-search algorithm is proportional to the size of the search space;
we process about 1k f-trees/second. The greedy heuristic is polynomial
in both $K$ and $L$, and in our scenario is 2-3 orders of magnitude
faster than full search.

{\noindent\bf Experiment 3: Query evaluation on flat data.}

We compared the performance of FDB, RDB, SQLite, and PostgreSQL for
query evaluation on flat input data. \nop{We measured the time needed
  to compute results without writing them back to disk. We also
  measured the size of the result as the number of its singletons; a
  singleton holds an 8 byte integer. \jakub{TODO: say someting about
    real sizes.}}  Figure~\ref{fig:exp3} shows the result sizes and
evaluation times for queries with up to four equality selections on
three ternary relations of increasing sizes in two settings: data
generated using a uniform distribution over the range $[1,100]$ (left
column) and using a more skewed, Zipf distribution (middle column).

The size gap between factorised and relational results is largest for
queries with fewer equality selections, since the results are larger
yet factorisable.  The plots support the claim that the sizes are
bounded by a power law, with a smaller exponent for FDB than for the
relational engines.

The rightmost column in Figure~\ref{fig:exp3} considers queries with
four relations, two binary relations of size $8^2 = 64$ and two
ternary relations of size $512 = 8^3$, whose values are drawn from
$[1,20]$ using uniform and Zipf distributions. This data\-set is
combinatorial in nature. Each equality selection in the query
decreases the number of result tuples by a constant factor of 20,
which is exhibited in the flat result size produced by RDB. FDB
factorises the up to 500M data values into less than 4k singletons for
all considered queries.

The execution time for all engines is approximately proportional to
their result sizes except for the millisecond region, where constant
overhead dominates. SQLite performs consistently slightly worse than
RDB, and PostgreSQL is about three times slower than SQLite. We used a
timeout of 100 seconds, which prohibited the relational engines to
complete in several cases (no plotted data points).

\nop{In summary, a combinatorial explosion of the number of tuples in the
query result benefits FDB: the tuples are processed one-by-one by RDB
(and the other relational engines), while FDB can represent them very
succinctly in factorised form.}


{\noindent\bf Experiment 4: Query evaluation on factorised
data.}

Figure~\ref{fig:exp4} compares the performance of FDB and RDB for
query evaluation on query results computed in Experiment 3 and with
f-plans computed in Experiment 2. The behaviour of SQLite and
PostgreSQL closely follows that of RDB.

FDB evaluates queries consisting of $L$ selections on factorised
representations. The quality of the resulting factorisation is
dictated by the quality of the f-plan. FDB uses the optimal f-plan
found by the full-search optimiser. Additional experiments (not
reported here) reveal that the f-plans found by the greedy optimiser
can be up to 50\% slower than the optimal f-plans.  This is a good
tradeoff, since the greedy optimiser runs fast even for large queries,
while the full-search optimiser explores an exponential space.

RDB just evaluates a selection with a conjunction of $L$ equality
conditions on the attributes of the input relation. This can be done
in one scan over the input relation. For FDB, the cost of the f-plan
may be non-trivial: the more the f-plan needs to unfold the
f-representation, the more expensive the evaluation becomes.
Figure~\ref{fig:exp4} suggests that FDB only unfolds the
f-representations to a small extent. Similar to query evaluation on
flat data, FDB shows up to 4 orders of magnitude improvement over RDB
for both evaluation time and result size. The gap closes once the size
of the input data decreases to about 1000 tuples and both FDB and RDB
perform in under 0.1 seconds.

Experiments 2 and 4 show that using f-representations for data
processing is sustainable in the sense that the quality of
factorisations, in particular their compactness and sizes, does not
decay with the number of operations on the data.

\bibliographystyle{abbrv}
\begin{small}
\bibliography{bibtex}
\end{small}

\end{document}